\documentclass[intlimits,twoside,a4paper]{article}

\usepackage[T2A]{fontenc} 
\usepackage[utf8]{inputenc}
\usepackage{cmpj3}
\usepackage{enumerate}
\usepackage{subfigure}
\usepackage{bm}

\issue{2021}{24}{4}{43701}
\doinumber{10.5488/CMP.24.43701}

\title{Phase diagrams of superconducting topological surface states}

\author[W.~H. Zhao \textsl{et al.}]{W.~H. Zhao, L.~L. Ding, B.~W. Zhou, J.~Y. Wu, Y. Bai, Z.~Y. Man, X. Luo\orcid{0000-0001-9124-6409}\thanks{Correspondence to: xiluo@usst.edu.cn}}
\address{Department of Physics, College of Science, University of Shanghai for Science and Technology, Shanghai 200093, PR China}

\date{Received January 17, 2021, in final form June 10, 2021}

\begin{document}
\maketitle
\begin{abstract}
In this paper, we present a detailed study on the phase diagrams of superconducting topological surface states, especially, focusing on the interplay between crystalline symmetry and topology of the effective BdG Hamiltonian. We show that for the $4\times 4$ kinematic Hamiltonian of the normal state, a mirror symmetry $\mathcal{M}$ can be defined, and for the $\mathcal{M}$-odd pairings, the classification of the $8\times8$ BdG Hamiltonian is $\mathbb{Z}\oplus\mathbb{Z}$, and the time-reversal symmetry is broken intrinsically. The topological non-trivial phase can support chiral Majorana edge modes, and can be realized in the thin films of iron-based superconductor such as FeSeTe.



\end{abstract}

\section{Introduction}

Since the proposal of Majorana fermion~\cite{majorana}, the search for it has been a long journey in condensed matter physics. The most famous proposal is the 2-dimensional $p$-wave superconductor~\cite{RG}, which is inspired by the Moore-Read state of in $\nu={5}/{2}$ fractional quantum Hall effect~\cite{MR}. Because of the nontrivial topology of bulk Chern number, there are chiral Majorana fermion edge modes and Majorana zero modes trapped in the vortices  in the 2-dimensional $p$-wave superconductor~\cite{MR}. Furthermore, these Majorana fermions can be described by a conformal field theory with the central charge $c={1}/{2}$, which is equivalent to the Ising model~\cite{cft}. Due to the non-abelian braiding of these Majorana fermions, a protocol for topological quantum computation based on them has been proposed~\cite{Kitaev,Ivanov,Georgiev,Nayak}. Because of the promising potential applications in realizing the topological quantum computation, lots of efforts have been poured into finding the $p$-wave superconductor, such as Sr$_2$RuO$_4$. Though the results are suggestive, they are not yet conclusive~\cite{p1,p2,p3}. There are other attempts besides finding the $p$-wave superconductor, among which, Fu and Kane's proposal is the most influential one~\cite{fukane}. Fu and Kane notice that the spectrum of an $s$-wave superconductor that couples the surface state of a strong topological insulator through the proximity effect is analogs to a spinless $p$-wave superconductor when the chemical potential is larger than the superconducting pairing order parameter, and they show the existence of Majorana zero modes trapped by vortices. Within their proposal, two ingredients are important, one is time-reversal symmetry, and the other is the spinless pairing. Fu and Kane's work has inspired lots of researches in finding topological superconductor and Majorana zero modes~\cite{been}, both theoretical and experimental. From the theoretical side, the time-reversal symmetry breaking is necessary to simulate an effective $p$-wave superconductor in Fu--Kane's scenario, which is usually achieved  by an external magnetic field, while the quantum anomalous Hall insulator provides another possibility~\cite{zhang1}. The quantum anomalous Hall insulator could be driven into a 2-dimensional chiral topological superconducting phase through the proximity effect of an $s$-wave superconductor, and a protocol is proposed for topological quantum computation based on the chiral Majorana edge modes~\cite{zhang2,zhang3}. Though this proposal remains controversial~\cite{wen,chang}, it is still  enlightening. Another theoretical progress is introducing the spin degree of freedom. Besides the usual spin-singlet pairing, the spin-triplet pairing has palyed an important role in topological superconductivity, such as Cu$_x$Bi$_2$Se$_3$ in 3D~\cite{fu2,CuBiSe} and bilayer Rashba System in 2D~\cite{nagaosa}. From the experimental side, mid‐gap bound states at zero bias voltage were found in hybrid superconductor-semiconductor nanowire devices which can be candidates for Majorana zero modes~\cite{mou}. Although other similar results were reported~\cite{eg1,eg2,eg3,eg4}, there can be other explanations by non-topological trivial bound states~\cite{exp}. Another possible Majorana zero mode was observed by spin selective Andreev reflection in the vortex of Bi$_2$Te$_3$/NbSe$_2$ heterostructure~\cite{jia}. Recently, Caroli--de Gennes--Matricon states in a vortex with a Majorana zero mode are found in iron-based superconductors, such as FeTe$_{0.55}$Se$_{0.45}$~\cite{lkong}, (Li$_{0.84}$Fe$_{0.16}$) OHFeSe~\cite{cchen}, LiFeAs~\cite{pzhang}, and CaKFe$_4$As$_4$~\cite{wliu}, which provides new promising platforms for quantum computation. The chiral Majorana edge modes in iron-based superconductors is also reported~\cite{zwang}, which suggests the existence of topological superconductivity in the iron-based superconductors. 

\begin{figure}[htb]
	\centerline{
	\begin{minipage}{0.23\textwidth}
		\centerline{\includegraphics[width=1\textwidth]{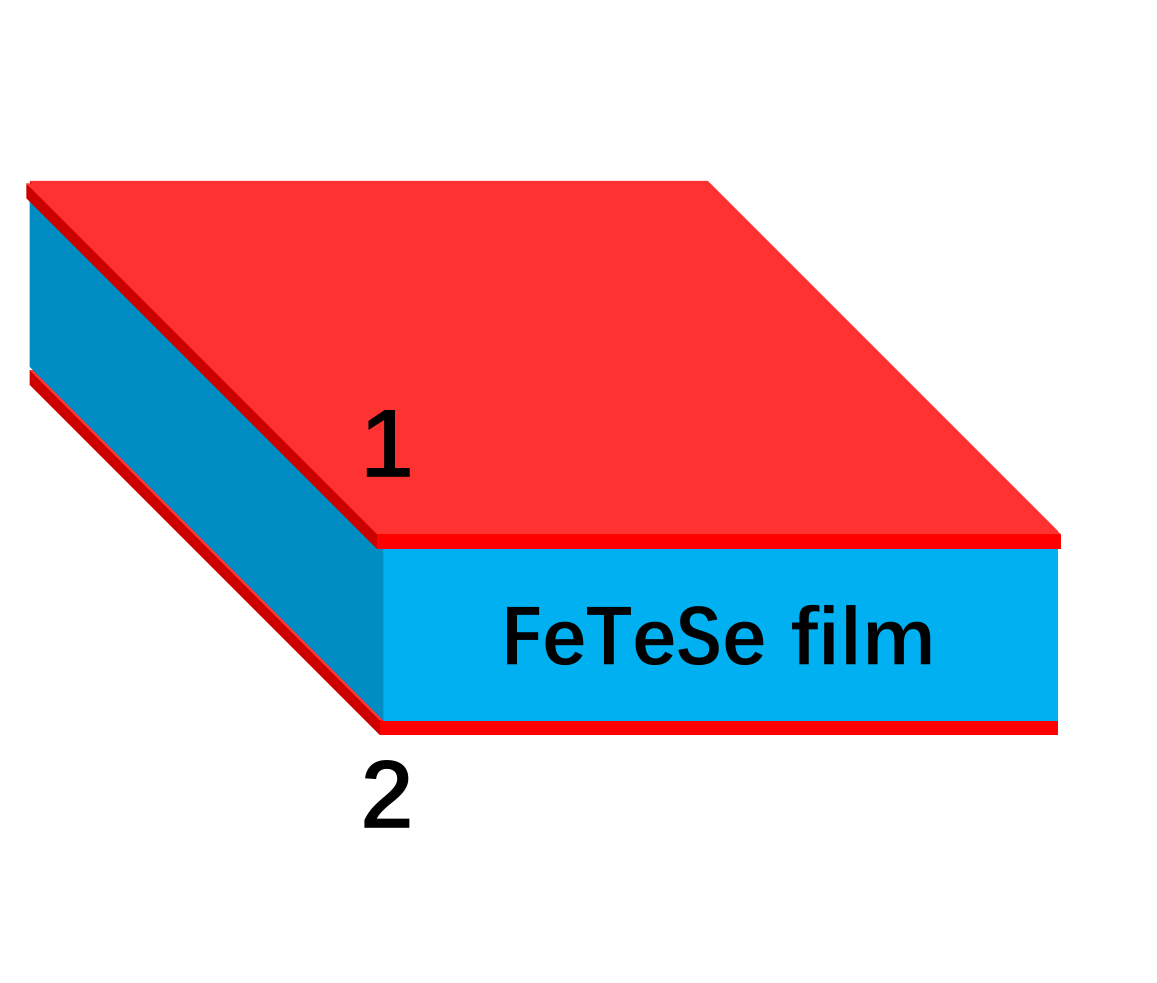}}
		\centerline{(a)}
	\end{minipage}
	\begin{minipage}{0.23\textwidth}
		\centerline{\includegraphics[width=1.1\textwidth]{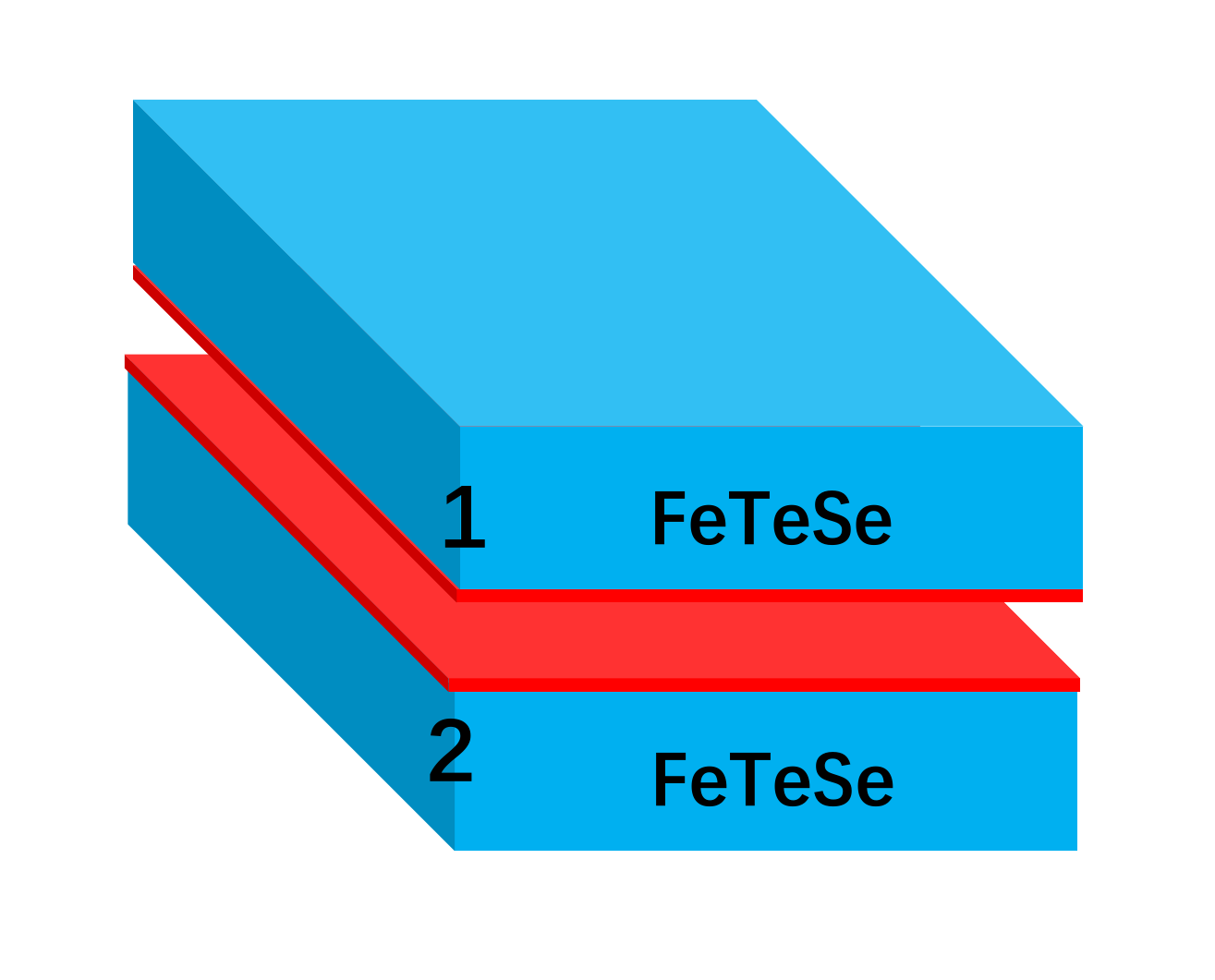}}
		\centerline{(b)}
	\end{minipage}
	\begin{minipage}{0.23\textwidth}
		\centerline{\includegraphics[width=1\textwidth]{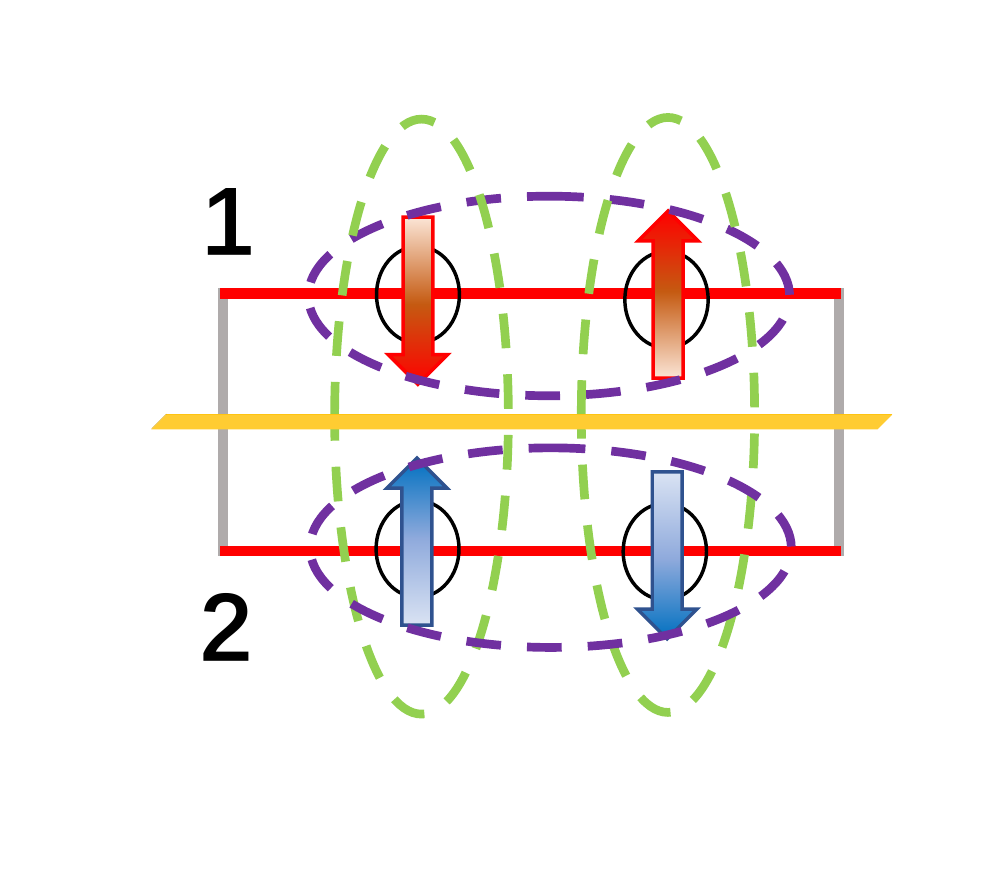}}
		\centerline{(c)}
	\end{minipage}}
	\caption{(Colour online)
		{The coupled superconducting topological surface states illustrated with FeTeSe. The red (blue) region stands for the surface (bulk). (a) Thin film with top ($1$) and bottom ($2$) superconducting topological surface states. (b) Opposing surfaces ($1$ and $2$) of two superconductors with topological surface states. (c) A schematic side view of the thin film in (a), the circles indicate Dirac fermions of the topological surface states, and the arrows are the spin degrees of freedom. The purple dashed circle stands for the intra-surface spin-singlet pairing, and the green dashed circle is for the inter-surface spin-triplet pairing. The orange line in the middle stands for the mirror plane. 
		}
		\label{figX}	}
\end{figure}

Inspired by these recent progresses, we have proposed a generic mechanism for realizing the topological chiral superconductor phase in superconducting topological surface states and a protocol for universal topological quantum computation based on the chiral Majorana edge modes in our previous works~\cite{luo1,luo2}. The key point is that, in Fu--Kane's scenario, the time-reversal symmetry breaking comes from the external magnetic field, which can create uncontrollable vortices in the superconductor that will drastically change the results of quantum computation~\cite{stern,bks}. Therefore, an intrinsic time-reversal symmetry breaking mechanism is needed. Furthermore, the time-reversal symmetry breaking in FeTeSe is also reported~\cite{jdr,zaki}, which hints at the existence of such an intrinsic mechanism. We notice that, for the topological surface states, the coexistence of inter-surface hopping, intra-surface spin-singlet pairing, and inter-surface spin-triplet pairing will spontaneously break time-reversal symmetry. This will be discussed more in detail in section 3. In this paper, we further study the phase diagram of superconducting topological surface states, and the interplay between crystalline mirror symmetry and the topology of the superconductor~\cite{ono,ms}. The intra-surface spin-singlet pairing and the inter-surface spin-triplet pairing are favored by the mirror symmetry. Here, the topological classification is a finer one, $\mathbb{Z}\oplus\mathbb{Z}$, due to the mirror symmetry~\cite{sato10,sato11,sato12,sato13}. Our model is possible to be realized in the thin films of iron-based superconductors (see figure~\ref{figX}). The reason is that the topological superconductivity in the  iron-based superconductors is believed to raise from the superconducting topological surface states, namely, the normal state in FeTeSe is a 3D topological insulator with the band inversion along the $\Gamma-Z$ direction~\cite{zjwang,gxu}. The topological surface states have opposite helicities on top and bottom surfaces, which provides an emergent mirror symmetry. The inter-surface spin-triplet pairings can be generated from a second order process of spin-orbital coupling (SOC) and spin-singlet pairings in the iron-based superconductor~\cite{yzhang1,yzhang2}. Furthermore, the growth of thin films of FeTeSe can be controlled by molecular beam epitaxy method or by exfoliation which are realizable by experimental techniques nowadays. The realization of such systems will provide a strong evidence for the chiral Majorana edge modes, which can provide a new route for realizing topological quantum computation~\cite{luo2}.

	The organization of this paper is as follows. In section~2, we describe our effective model Bogoliubov--de~Gennes (BdG) Hamiltonian. We present the symmetry analysis of the BdG Hamiltonian in section~3. The BdG Hamiltonian can be block diagonalized due to the mirror symmetry, and for the mirror-odd pairings, namely, the intra-surface spin-singlet pairing and the inter-surface spin-triplet pairing, the block diagonalized Hamiltonian becomes a direct sum of two effective $p$-wave superconductors. We discuss the topological classification in section~4, which is confirmed by the phase diagram and lattice model study in section~5. In section~6, we discuss the effects of the relative phase between the paring strengths, present the topological classification of mirror protected superconductor. The last section is devoted to conclusions.

\section{Model Hamiltonian}

We choose the Nambu basis as $\Psi_N=(\psi_{1\uparrow \textbf{q}},\psi_{1\downarrow \textbf{q}},\psi_{2 \uparrow \textbf{q}},\psi_{2 \downarrow \textbf{q}},\psi^{\dag}_{1\uparrow -\textbf{q}},\psi^{\dag}_{1\downarrow -\textbf{q}},\psi^{\dag}_{2\uparrow -\textbf{q}},\psi^{\dag}_{2 \downarrow -\textbf{q}})^T$, with the BdG Hamiltonian (the Fermi velocity $v_{\rm F}$ is set to unity),
\begin{equation}
h=\left(
\begin{array}{cc}
\tilde{h}_0(\textbf{q})_{4\times4} & \tilde{\Delta}_{4\times4}^\dag \\
\tilde{\Delta}_{4\times4} & -\tilde{h}^*_0(-\textbf{q})_{4\times4} \\
\end{array}
\right), \label{h}
\end{equation}
\begin{eqnarray}
\tilde{h}_0(\textbf{q})_{4\times4}
&=&
\setlength{\arraycolsep}{0.1pt}
\textbf{q}\cdot {\boldsymbol\sigma}\chi_z-\mu\sigma_0\chi_0 +t\sigma_0\chi_x+\lambda\sigma_z\chi_0\nonumber\\
&=&
\setlength{\arraycolsep}{0.1pt}
\left(
\begin{array}{cccc}
	\lambda -\mu  & q_1-\ri q_2 & t & 0 \\
	q_1+\ri q_2 & -\lambda -\mu  & 0 & t \\
	t & 0 & \lambda -\mu  & -q_1+\ri q_2 \\
	0 & t & -q_1-\ri q_2 & -\lambda -\mu  \\
\end{array} \label{h0}
\right),\\
\tilde{\Delta}_{4\times4}&=& \ri\Delta\sigma_y\chi_z-\ri\Delta_s\sigma_y\chi_x-\ri\Delta_t\sigma_x\chi_y \nonumber\\
&=&
\setlength{\arraycolsep}{0.1pt}
\left(
\begin{array}{cccc}
0 & \Delta  & 0 & -\Delta _s-\Delta _t \\
-\Delta  & 0 & \Delta _s-\Delta _t & 0 \\
0 & -\Delta _s+\Delta _t & 0 & -\Delta  \\
\Delta _s+\Delta _t & 0 & \Delta  & 0 \\
\end{array}
\right),
\end{eqnarray}
where 1 and 2 are the surface indices (see figure~\ref{figX}),   $\uparrow$ and $\downarrow$ label the
pseudospin states in the presence of SOC, and $\textbf{q}$ labels the momentum. $\chi_\mu$ and $\sigma_\mu$ act on the surface and spin indices, respectively. The Hamiltonian $\tilde{h}_0(\textbf{q})$ describes two coupled topological surface states with opposite helicities and the inter-surface hopping constant $t$, which can be realized on the surfaces of the thin film of a 3-dimensional topological insulator. $\mu$ is the chemical potential,  and  $\lambda$ is the Zeeman term which can be set to zero in our model Hamiltonian. $\Delta$ is the spin singlet intra-surface pairing strength, and $\Delta_s$ and $\Delta_t$ are the spin singlet (surface triplet) and triplet (surface singlet) inter-surface pairing constants [see figure~\ref{figX}~(c)].  As described in Introduction, this effective model can be realized in thin films of iron-based superconductors near the 2D $\Gamma$ point. For the case when $\mu=0$ and $\lambda=0$, the spectra are
\begin{equation}
	E_{\kappa\eta\xi}=\kappa\sqrt{{\bf q}^2+2\eta \sqrt{\Delta_t^2 \left[\Delta_s^2+(t+\xi\Delta )^2\right]+\Delta_s^2 \left({\bf q}^2\right)}+\Delta_s^2+\Delta_t^2+(t+\xi\Delta )^2}, \label{E}
\end{equation} 
with ${\bf q}^2=q_1^2+q_2^2$ and $\kappa,\eta,\xi=\pm$. The topological phase transition occurs when the gap closes at the $\Gamma$ point. Therefore, the phase boundary in the parameter space of $t$, $\Delta$, $\Delta_t$, and $\Delta_s$ ($\mu$, $\lambda$ can also be included) is determined by the zero lines of $E_{\kappa\eta\xi}({\bf q}=0)$. For example, when $\Delta_s=0$, the above spectra become,
\begin{equation}
	E_{\kappa\eta\xi}(\Delta_s=0)=\kappa\sqrt{{\bf q}^2+(t+\eta\Delta+\xi\Delta)^2},
\end{equation}
which share the same form as Dirac fermions, and the Chern number is determined by the signs of the corresponding mass terms. More details on the topological classification will be discussed in section~5. 

\section{Symmetry analysis}

\subsection{Particle-hole symmetry}
The BdG Hamiltonian~$(\ref{h})$ has a particle-hole symmetry  $\mathcal{C}h(\textbf{q})\mathcal{C}^{-1}=-h(-\textbf{q})$ with $\mathcal{C}=\sigma_0\chi_0\tau_x\mathcal{K}$, where $\tau_i$ is the Pauli matrix that acts on the particle-hole space, and  $\mathcal{K}$ is the complex conjugation.

\subsection{Inversion symmetry}
For the normal state Hamiltonian $\tilde{h}_0(\textbf{q})_{4\times4}$~(\ref{h0}), we introduce the inversion operator $\mathcal{I}$ satisfying $\mathcal{I} \tilde{h}_0(\textbf{q})_{4\times4} \mathcal{I}^{-1}=\tilde{h}_0(-\textbf{q})_{4\times4}$. The non-trivial choice of $\mathcal{I}$ is
\begin{equation}
	\mathcal{I}=\sigma_0\chi_x.
\end{equation}
Under $\mathcal{I}$, 
\begin{equation}
	\mathcal{I}\tilde{\Delta}_{4\times4}\mathcal{I}^{-1}=-\ri\Delta\sigma_y\chi_z-\ri\Delta_s\sigma_y\chi_x+\ri\Delta_t\sigma_x\chi_y,
\end{equation}
namely, $\Delta$ and $\Delta_t$ terms are $\mathcal{I}$-{odd} while $\Delta_s$ term is $\mathcal{I}$-{even}.

The inversion symmetry $\mathcal{I}$ can be generalized to the BdG Hamiltonian~(\ref{h}) by defining
\begin{equation}
	\mathcal{P}_\pm=
	\left(
	\begin{array}{cccc}
		\mathcal{I} & 0\\
		0 & \pm\mathcal{I} \\
	\end{array}
	\right),
\end{equation}
namely, $\mathcal{P}_+=\sigma_0\chi_x\tau_0$, $\mathcal{P}_-=\sigma_0\chi_x\tau_z$, and {$\{\mathcal{P}_-,\mathcal{C}\}=0$, $[\mathcal{P}_+,\mathcal{C}]=0$.}

\subsection{Mirror symmetry}

For the normal state Hamiltonian $\tilde{h}_0(\textbf{q})_{4\times4}$~(\ref{h0}), there is an emergent mirror symmetry $\mathcal{M}$ with respect to $xy$ plane, namely, $\mathcal{M} \tilde{h}_0(\textbf{q})_{4\times4} \mathcal{M}^{-1}=\tilde{h}_0(\textbf{q})_{4\times4}$ with
\begin{equation}
\mathcal{M}=-\ri\sigma_z\chi_x, \quad \mathcal{M}^2=-1.
\end{equation}
This mirror symmetry $\mathcal{M}$ originates from the helical structure of the topological surface states on the two surfaces. Under $\mathcal{M}$, the pairing terms transform as 
\begin{equation}
\mathcal{M}\tilde{\Delta}_{4\times4}\mathcal{M}^T=
-\ri\Delta\sigma_y\chi_z-\ri\Delta_s\sigma_y\chi_x+\ri\Delta_t\sigma_x\chi_y,
\end{equation}
in other words, $\Delta$ and $\Delta_t$ terms are $\mathcal{M}$-{odd} while $\Delta_s$ term is $\mathcal{M}$-{even}. 

The mirror symmetry $\mathcal{M}$ can also be generalized to the BdG Hamiltonian~(\ref{h}) by defining
\begin{equation}
\mathcal{M}_{\pm}=
\left(
\begin{array}{cccc}
\mathcal{M} & 0\\
0 & \pm\mathcal{M}^* \\
\end{array}
\right),
\end{equation}
namely, $\mathcal{M}_+=-\ri\sigma_z\chi_x\tau_z$, $\mathcal{M}_-=-\ri\sigma_z\chi_x\tau_0$, and {$[ \mathcal{M}_+,\mathcal{C}]=0$, $\{\mathcal{M}_-,\mathcal{C}\}=0$.}  $\mathcal{P}_\pm$ commute with $\mathcal{M}_\pm$, and they can be diagonalized simultaneously.

The anti-commutation relation $\{\mathcal{M}_-,\mathcal{C} \}=0$ ensures that the BdG Hamiltonian~(\ref{h}) can be block diagonalized in a basis that diagonalizes $\mathcal{M}_-$, and each block has a particle-hole symmetry~\cite{sato10,sato11,sato12,sato13}. More details on the topological classification are presented in section~4. Therefore, let us first focus on $\mathcal{M}_-$. Consider the following unitary transformation that diagonalizes $\mathcal{M}_-$, 
\begin{equation}
U_-=\frac{1}{\sqrt{2}}
\left(
\begin{array}{cccccccc}
0 & 0 & 0 & 0 & 1 & 0 & 1 & 0 \\
0 & 1 & 0 & -1 & 0 & 0 & 0 & 0 \\
1 & 0 & 1 & 0 & 0 & 0 & 0 & 0 \\
0 & 0 & 0 & 0 & 0 & 1 & 0 & -1 \\
1 & 0 & -1 & 0 & 0 & 0 & 0 & 0 \\
0 & 0 & 0 & 0 & 0 & 1 & 0 & 1 \\
0 & 0 & 0 & 0 & 1 & 0 & -1 & 0 \\
0 & 1 & 0 & 1 & 0 & 0 & 0 & 0 \\
\end{array}
\right),
\end{equation}
then 
\begin{eqnarray}
&&U_-\mathcal{P}_+U_-^{-1}=\rm{diag}(1,-1,1,-1,-1,1,-1,1),\\ &&U_-\mathcal{P}_-U_-^{-1}=\rm{diag}(-1,-1,1,1,-1,-1,1,1),\\ &&U_-\mathcal{M}_+U_-^{-1}=\rm{diag}(\ri,-\ri,-\ri,\ri,\ri,-\ri,-\ri,\ri), \\ &&U_-\mathcal{M}_-U_-^{-1}=\rm{diag}(-\ri,-\ri,-\ri,-\ri,\ri,\ri,\ri,\ri),
\end{eqnarray}
 and the model Hamiltonian becomes
\begin{small}
\begin{eqnarray}
\setlength{\arraycolsep}{0.01pt}
&&U_- h(\textbf{q})U_-^{-1}\nonumber\\
&=&
\left(
\begin{array}{cccccccc}
-t-\lambda +\mu  & \Delta +\Delta _t & 0 & q_1+\ri q_2 & 0 & 0 & 0 & -\Delta _s \\
\Delta +\Delta _t & -t-\lambda -\mu  & q_1+\ri q_2 & 0 & 0 & 0 & \Delta _s & 0 \\
0 & q_1-\ri q_2 & t+\lambda -\mu  & -\Delta -\Delta _t & 0 & \Delta _s & 0 & 0 \\
q_1-\ri q_2 & 0 & -\Delta -\Delta _t & t+\lambda +\mu  & -\Delta _s & 0 & 0 & 0 \\
0 & 0 & 0 & -\Delta _s & -t+\lambda -\mu  & -\Delta +\Delta _t & 0 & q_1-\ri q_2 \\
0 & 0 & \Delta _s & 0 & -\Delta +\Delta _t & -t+\lambda +\mu  & q_1-\ri q_2 & 0 \\
0 & \Delta _s & 0 & 0 & 0 & q_1+\ri q_2 & t-\lambda +\mu  & \Delta -\Delta _t \\
-\Delta _s & 0 & 0 & 0 & q_1+\ri q_2 & 0 & \Delta -\Delta _t & t-\lambda -\mu  \\
\end{array}
\right),\nonumber\\
\end{eqnarray}
\end{small}
with the basis $U_-\Psi_N=({1}/{\sqrt{2}})
(\psi^\dag_{1\uparrow -\textbf{q}}+\psi^\dag_{2\uparrow -\textbf{q}},\psi_{1\downarrow \textbf{q}}-\psi_{2\downarrow \textbf{q}},\psi_{1\uparrow \textbf{q}}+\psi_{2\uparrow \textbf{q}}, \psi^\dag_{1\downarrow -\textbf{q}}-\psi^\dag_{2\downarrow -\textbf{q}},\psi_{1\uparrow \textbf{q}}-\psi_{2\uparrow \textbf{q}}, \psi^\dag_{1\downarrow -\textbf{q}}+\psi^\dag_{2\downarrow -\textbf{q}},\psi^\dag_{1\uparrow -\textbf{q}}-\psi^\dag_{2\uparrow -\textbf{q}},\psi_{1\downarrow \textbf{q}}+\psi_{2\downarrow \textbf{q}})^T$. Clearly, the existence of $\Delta_s$ breaks the $\mathcal{M}_-$ symmetry. When $\Delta_s=0$, the above Hamiltonian becomes block diagonal $h_1\oplus h_2$ with
\begin{eqnarray}
	h_1(\textbf{q})&=& -(t+\lambda)\sigma_0\Gamma_z +\mu\sigma_z\Gamma_z +(\Delta+\Delta_t)\sigma_x\Gamma_z+q_1\sigma_x\Gamma_x-q_2\sigma_x\Gamma_y \nonumber\\
	&=&
	\left(
\begin{array}{cccc}
-t-\lambda +\mu  & \Delta +\Delta _t & 0 & q_1+\ri q_2 \\
\Delta +\Delta _t & -t-\lambda -\mu  & q_1+\ri q_2 & 0 \\
0 & q_1-\ri q_2 & t+\lambda -\mu  & -\Delta -\Delta _t \\
q_1-\ri q_2 & 0 & -\Delta -\Delta _t & t+\lambda +\mu  \\
\end{array}
	\right),\\
	h_2(\textbf{q})&=& -(t-\lambda)\sigma_0\Gamma_z -\mu\sigma_z\Gamma_z -(\Delta-\Delta_t)\sigma_x\Gamma_z+q_1\sigma_x\Gamma_x+q_2\sigma_x\Gamma_y \nonumber\\
	&=&
	\left(
	\begin{array}{cccc}
	-t+\lambda -\mu  & -\Delta +\Delta _t & 0 & q_1-\ri q_2 \\
	-\Delta +\Delta _t & -t+\lambda +\mu  & q_1-\ri q_2 & 0 \\
	0 & q_1+\ri q_2 & t-\lambda +\mu  & \Delta -\Delta _t \\
	q_1+\ri q_2 & 0 & \Delta -\Delta _t & t-\lambda -\mu  \\
	\end{array}
	\right),
\end{eqnarray}
where $\sigma_\mu$ acts on the spin space and $\Gamma_\mu$ acts on the parity space defined by $P_-$, which mixes the surface and particle-hole indices. The spectra for $h_1$ and $h_2$ read
\begin{eqnarray}
	E_{1\kappa\eta}&=&\kappa\sqrt{{\bf q}^2+2\eta  \sqrt{(\lambda +t)^2 \left[(\Delta_t+\Delta )^2+\mu ^2\right]+\mu ^2 {\bf q}^2}+(\Delta_t+\Delta )^2+\mu ^2+(\lambda +t)^2},
\end{eqnarray}
\begin{eqnarray}
		E_{2\kappa\eta}&=&\kappa\sqrt{{\bf q}^2+2\eta  \sqrt{(\lambda -t)^2 \left[(\Delta_t-\Delta )^2+\mu ^2\right]+\mu ^2 {\bf q}^2}+(\Delta_t-\Delta )^2+\mu ^2+(\lambda -t)^2}.
\end{eqnarray}
These spectra are consistent with the ones in~(\ref{E}) for $\Delta_s=0$.

Because $\{\mathcal{M}_-,\mathcal{C}\}=0$, for each $h_1$ and $h_2$, a particle-hole operator $\tilde{\mathcal{C}}_-$ can be defined~\cite{sato10},

\begin{equation}
\tilde{\mathcal{C}}_-=\sigma_0\Gamma_x\mathcal{K}, \quad \tilde{\mathcal{C}}_-^2=1,
\end{equation}
such that $\tilde{\mathcal{C}}_-h_1(\textbf{q})\tilde{\mathcal{C}}_-^{-1}=-h_1(-\textbf{q})$, and $\tilde{\mathcal{C}}_-h_2(\textbf{q})\tilde{\mathcal{C}}_-^{-1}=-h_2(-\textbf{q})$, namely, each of  $h_1$ and $h_2$ describes a BdG Hamiltonian, and the edge modes are chiral Majorana fermions. 

Obviously, $h_1(\Delta,\lambda, \mu, q_2)=h_2(-\Delta,-\lambda, -\mu,  -q_2)$. Without the Zeeman field $\lambda=0$, then 
$(\sigma_x\Gamma_z\mathcal{K})h_1(\textbf{q})(\sigma_x\Gamma_z\mathcal{K})^{-1}=h_2(-\textbf{q})$ for $\Delta=0$, and  $(\ri\sigma_y\Gamma_0\mathcal{K})h_1(\textbf{q})(\ri\sigma_y\Gamma_0\mathcal{K})^{-1}=h_2(-\textbf{q})$ for $\Delta_t=0$. These two equations indicate the existence of time reversal symmetry, which will be discussed in the next subsection.

If we denote $U_-\Psi_N=(\Psi_1,\Psi_2)^T$ with $\Psi_1=\left(a_\uparrow,a_\downarrow,a_\uparrow^\dag,a_\downarrow^\dag\right)^T$, and $\Psi_2=\left(b_\uparrow,b_\downarrow,b_\uparrow^\dag,b_\downarrow^\dag\right)^T$. These renamed annihilation and creation operators satisfy the anti-commutation relation of fermions. Then,
\begin{eqnarray}
	\sum_\textbf{q}\Psi_1^\dag h_1 \Psi_1&=&\sum_\textbf{q}\left[(-t+\mu)a^\dag_{\uparrow \textbf{q}}a_{\uparrow \textbf{q}}+(-t-\mu)a^\dag_{\downarrow \textbf{q}}a_{\downarrow \textbf{q}}\right.\nonumber\\
	&&+(\Delta+\Delta_t)\left(a^\dag_{\uparrow \textbf{q}}a_{\downarrow \textbf{q}}+a^\dag_{\downarrow \textbf{q}}a_{\uparrow \textbf{q}}\right)\nonumber\\
	&&+\left. (q_1+\ri q_2)\left(a^\dag_{\uparrow \textbf{q}}a^\dag_{\downarrow -\textbf{q}}+a^\dag_{\downarrow \textbf{q}}a^\dag_{\uparrow -\textbf{q}}\right)+h.c.\right],\label{p+}\\
	\sum_\textbf{q}\Psi_2^\dag h_2 \Psi_2&=&\sum_\textbf{q}\left[(-t-\mu)b^\dag_{\uparrow \textbf{q}}b_{\uparrow \textbf{q}}+(-t+\mu)b^\dag_{\downarrow \textbf{q}}b_{\downarrow \textbf{q}}\right.\nonumber\\
	&&+(-\Delta+\Delta_t)\left(b^\dag_{\uparrow \textbf{q}}b_{\downarrow \textbf{q}}+b^\dag_{\downarrow \textbf{q}}b_{\uparrow \textbf{q}}\right)\nonumber\\
	&&+\left. (q_1-\ri q_2)\left(b^\dag_{\uparrow \textbf{q}}b^\dag_{\downarrow -\textbf{q}}+b^\dag_{\downarrow \textbf{q}}b^\dag_{\uparrow -\textbf{q}}\right)+h.c.\right],\label{p-}
\end{eqnarray}
which represent two independent $p$-wave spin triplet superconductors when none of $t$, $\Delta$, and $\Delta_t$ is zero. 

Now let us focus on $\mathcal{M}_+$. Consider the unitary transformation $U_+$ that diagonalizes $\mathcal{M}_+$,
\begin{equation}
U_+=\frac{1}{\sqrt{2}}
\left(
\begin{array}{cccccccc}
	0 & 0 & 0 & 0 & 1 & 0 & -1 & 0 \\
	0 & 1 & 0 & -1 & 0 & 0 & 0 & 0 \\
	1 & 0 & 1 & 0 & 0 & 0 & 0 & 0 \\
	0 & 0 & 0 & 0 & 0 & 1 & 0 & 1 \\
	1 & 0 & -1 & 0 & 0 & 0 & 0 & 0 \\
	0 & 0 & 0 & 0 & 0 & 1 & 0 & -1 \\
	0 & 0 & 0 & 0 & 1 & 0 & 1 & 0 \\
	0 & 1 & 0 & 1 & 0 & 0 & 0 & 0 \\
\end{array}
\right),
\end{equation}
then 
\begin{eqnarray}
	U_+\mathcal{P}_+U_+^{-1}=\rm{diag}(-1,-1,1,1,-1,-1,1,1),\\
	U_+\mathcal{P}_-U_+^{-1}=\rm{diag}(1,-1,1,-1,-1,1,-1,1),\\
	U_+\mathcal{M}_+U_+^{-1}=\rm{diag}(-\ri,-\ri,-\ri,-\ri,\ri,\ri,\ri,\ri),\\
	U_+\mathcal{M}_-U_+^{-1}=\rm{diag}(\ri,-\ri,-\ri,\ri,\ri,-\ri,-\ri,\ri),
\end{eqnarray}
and the model Hamiltonian becomes
\begin{small}
\begin{eqnarray}
&&U_+h(\textbf{q})U_+^{-1}\nonumber\\
&=&
\setlength{\arraycolsep}{2pt}
\left(
\begin{array}{cccccccc}
	t-\lambda +\mu  & \Delta _s & 0 & q_1+\ri q_2 & 0 & 0 & 0 & \Delta -\Delta _t \\
	\Delta _s & -t-\lambda -\mu  & q_1+\ri q_2 & 0 & 0 & 0 & \Delta +\Delta _t & 0 \\
	0 & q_1-\ri q_2 & t+\lambda -\mu  & \Delta _s & 0 & -\Delta -\Delta _t & 0 & 0 \\
	q_1-\ri q_2 & 0 & \Delta _s & -t+\lambda +\mu  & -\Delta +\Delta _t & 0 & 0 & 0 \\
	0 & 0 & 0 & -\Delta +\Delta _t & -t+\lambda -\mu  & -\Delta _s & 0 & q_1-\ri q_2 \\
	0 & 0 & -\Delta -\Delta _t & 0 & -\Delta _s & t+\lambda +\mu  & q_1-\ri q_2 & 0 \\
	0 & \Delta +\Delta _t & 0 & 0 & 0 & q_1+\ri q_2 & -t-\lambda +\mu  & -\Delta _s \\
	\Delta -\Delta _t & 0 & 0 & 0 & q_1+\ri q_2 & 0 & -\Delta _s & t-\lambda -\mu  \\
\end{array}
\right),
\end{eqnarray}
\end{small}
with the basis  $U_+\Psi_N=\frac{1}{\sqrt{2}}
(\psi^\dag_{1\uparrow -\textbf{q}}-\psi^\dag_{2\uparrow -\textbf{q}},\psi_{1\downarrow \textbf{q}}-\psi_{2\downarrow \textbf{q}},\psi_{1\uparrow \textbf{q}}+\psi_{2\uparrow \textbf{q}}, \psi^\dag_{1\downarrow -\textbf{q}}+\psi^\dag_{2\downarrow -\textbf{q}},\psi_{1\uparrow \textbf{q}}-\psi_{2\uparrow \textbf{q}}, \psi^\dag_{1\downarrow -\textbf{q}}-\psi^\dag_{2\downarrow -\textbf{q}},\psi^\dag_{1\uparrow -\textbf{q}}+\psi^\dag_{2\uparrow -\textbf{q}},\psi_{1\downarrow \textbf{q}}+\psi_{2\downarrow \textbf{q}})^T$. Clearly, the existence of $\Delta$ and $\Delta_t$ break the $M_+$ symmetry. When $\Delta=\Delta_t=0$, the above Hamiltonian becomes block diagonal $h_3\oplus h_4$ with
\begin{eqnarray}
h_3(\textbf{q})&=& t\sigma_z\Gamma_0+\mu\sigma_z\Gamma_z-\lambda\sigma_0\Gamma_z+\Delta_s\sigma_x\Gamma_0+q_1\sigma_x\Gamma_x-q_2\sigma_x\Gamma_y \nonumber\\
&=&\left(
\begin{array}{cccc}
	t-\lambda +\mu  & \Delta _s & 0 & q_1+\ri q_2 \\
	\Delta _s & -t-\lambda -\mu  & q_1+\ri q_2 & 0 \\
	0 & q_1-\ri q_2 & t+\lambda -\mu  & \Delta _s \\
	q_1-\ri q_2 & 0 & \Delta _s & -t+\lambda +\mu
\end{array}
\right),\\
h_4(\textbf{q}) &=& -t\sigma_z\Gamma_0-\mu\sigma_z\Gamma_z+\lambda\sigma_0\Gamma_z-\Delta_s\sigma_x\Gamma_0+q_1\sigma_x\Gamma_x+q_2\sigma_x\Gamma_y \nonumber\\
&=&
\left(
\begin{array}{cccc}
	-t+\lambda -\mu  & -\Delta _s & 0 & q_1-\ri q_2 \\
	-\Delta _s & t+\lambda +\mu  & q_1-\ri q_2 & 0 \\
	0 & q_1+\ri q_2 & -t-\lambda +\mu  & -\Delta _s \\
	q_1+\ri q_2 & 0 & -\Delta _s & t-\lambda -\mu  \\
\end{array}
\right),
\end{eqnarray}
where $\Gamma_\mu$ acts on the parity space defined by the  $\mathcal{P}_+$. The spectra for $h_3$ and $h_4$ with $\lambda=0$ read
\begin{eqnarray}
	E_{3\kappa\eta}=E_{4\kappa\eta}=\kappa\sqrt{{\bf q}^2+2\eta \sqrt{b^2 \left({\bf q}^2\right)+\mu ^2 \left({\bf q}^2+t^2\right)}+b^2+\mu ^2+t^2}.
\end{eqnarray}
These spectra are consistent with the ones in~(\ref{E}) for $\Delta=\Delta_t=0$. The degeneracy of the spectra comes from time-reversal symmetry.

Since $[\mathcal{M}_+,\mathcal{C}]=0$, for each $h_3$ and $h_4$, the particle-hole symmetry does not exist~\cite{sato10}. Therefore, the edge modes of $h_3(h_4)$ are charged fermions instead of Majorana fermions. $(\ri\sigma_y\Gamma_0\mathcal{K})h_3(\textbf{q})(\ri\sigma_y\Gamma_0\mathcal{K})^{-1}=h_4(-\textbf{q})$ indicates a time reversal symmetry.  

\subsection{(Modified) time-reversal symmetry}

In this subsection, we choose $\lambda=0$ since the Zeeman term always breaks the time-reversal symmetry.

Each of $h_1$ and $h_2$ breaks the time-reversal symmetry when $t\neq 0$. When $t=0$, $(\ri\sigma_y\Gamma_x\mathcal{K})h_{1,2}(\textbf{q})(\ri\sigma_y\Gamma_x\mathcal{K})^{-1}=h_{1,2}(-\textbf{q})$, namely, each of $h_1$ and $h_2$ is time-reversal symmetric. As mentioned in the above subsection, when $\Delta=0$, there is a time-reversal symmetry for $U_-hU_-^{-1}$ with $\mathcal{T}_{\Delta=0}=\ri\sigma_x\Gamma_z\Theta_y\mathcal{K}$, and one for $\Delta_t=0$ with $\mathcal{T}_{\Delta_t=0}=\ri\sigma_y\Gamma_0\Theta_x\mathcal{K}$, where $\Theta_y$ acts on the two dimensional space spanned by $h_1\oplus h_2$ and $\mathcal{T}^2=-1$. We summarize the results in table~\ref{tab:1}. From table~\ref{tab:1}, there is no common time-reversal operator shared by $\Delta$-, $t$-, and $\Delta_t$-terms, in other words, the time-reversal symmetry is spontaneously broken. 

\begin{table}[htb]
\caption{Summary of the time-reversal operators for Hamiltonian~(\ref{h}) in the untransformed Nambu basis. $\pm$ represent the corresponding term is time-reversal invariant or not.}
\label{tab:1}	
\begin{center}
	\begin{tabular}{|c|c|c|c|c|c|c|c|}
		\hline $\mathcal{T}$ & $\mathcal{T}^2$ & $\mu$-term & $\Delta$-term& $t$-term & $\Delta_t$-term & $\lambda$-term \\
		\hline $\ri\sigma_y\chi_0\tau_0\mathcal{K}$ & $-1$ & $+$ & $+$ & $+$ & $-$ & $-$ \\
		\hline $\ri\sigma_y\chi_0 \tau_z\mathcal{K}$ & $-1$  & $+$ & $-$ & $+$ & $+$ & $-$ \\
		\hline $ \ri\sigma_y\chi_z\tau_0\mathcal{K}$ & $-1$  & $+$ & $+$ & $-$ & $+$ & $-$ \\
		\hline
	\end{tabular}
	\end{center}
\end{table}

Similarly, for $h_3$ and $h_4$, $U_+hU_+^{-1}$ is also time-reversal symmetric with $\mathcal{T}_-=\ri\sigma_y\Gamma_0\Theta_x\mathcal{K}$.

\section{Topological classification}

Though the topological classification of mirror protected superconductor has been discussed in~\cite{sato10,sato11,sato12,sato13}, our model Hamiltonian provides a vivid example of mirror protected crystalline superconductor with an \emph{intrinsic} time-reversal breaking chiral phase with $q$-\emph{independent} pairings. Without time-reversal symmetry, the {$\mathcal{M}$-odd} pairings provide a $\mathbb{Z}\oplus\mathbb{Z}$ topological structure because each of $h_1$ and $h_2$ belongs to Class D and provides a $\mathbb{Z}$~\cite{10-fold}. The detailed calculations on Chern numbers are presented in the next section. With time-reversal symmetry, for both $\mathcal{M}$-even and -odd cases, they are classified by $\mathbb{Z}_2$~\cite{10-fold}.

\section{Phase diagrams and numerics}

\begin{figure}
\centerline{
	\begin{minipage}[b]{0.23\textwidth}
		\subfigure[]{
			\includegraphics[width=1\textwidth]{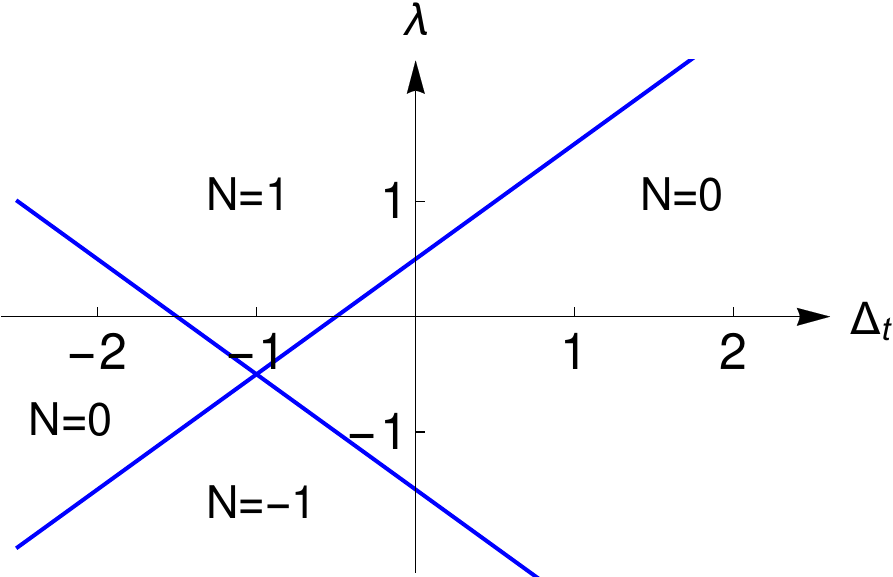}
		}			
		\subfigure[]{
			\includegraphics[width=1\textwidth]{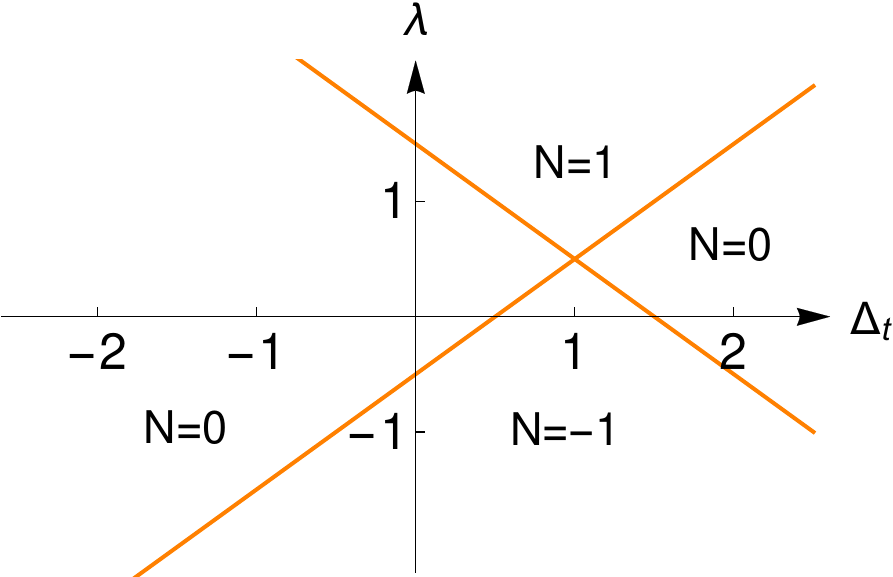}
		}   
	\end{minipage}
	\begin{minipage}[b]{0.23\textwidth}
		\subfigure[]{
			\includegraphics[width=1\textwidth]{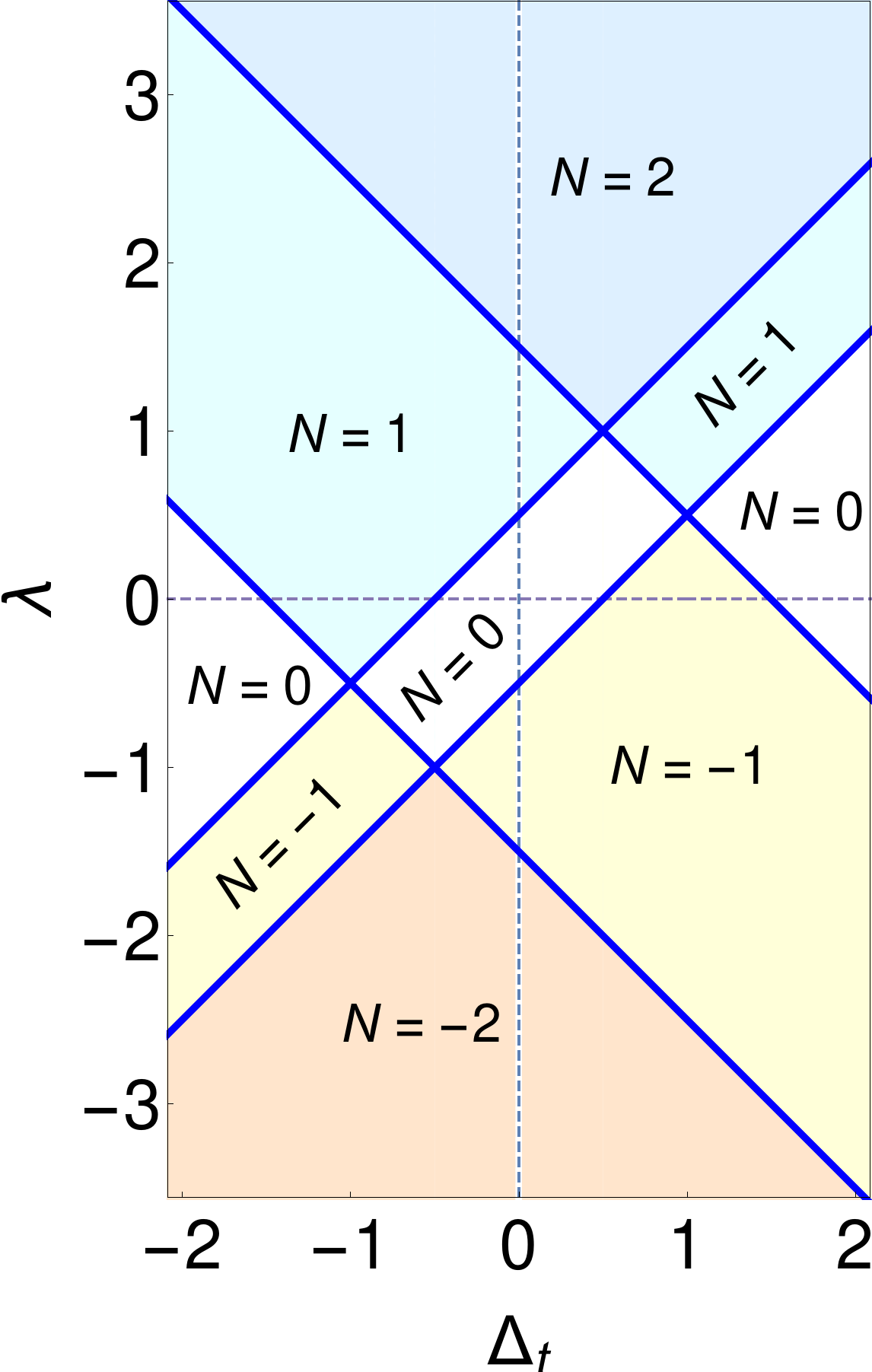}
		}
	\end{minipage}	}
	\caption{(Colour online)
		Topological phase diagrams of (a) $h_1$, (b) $h_2$, (c) $h$ in the $\Delta_t$-$\lambda$ plane for $\mu=0$, $\Delta_s=0$, $\Delta=1$, and $t=0.5$. The dashed lines in (c) stand for $\Delta_t=0$ and $\lambda=0$. $N$ is the total Chern number, and the solid lines are the phase boundaries.}
	\label{fig1}
\end{figure}

\begin{figure}
	\centerline{
	\begin{minipage}{0.23\textwidth}
		\centerline{\includegraphics[width=1\textwidth]{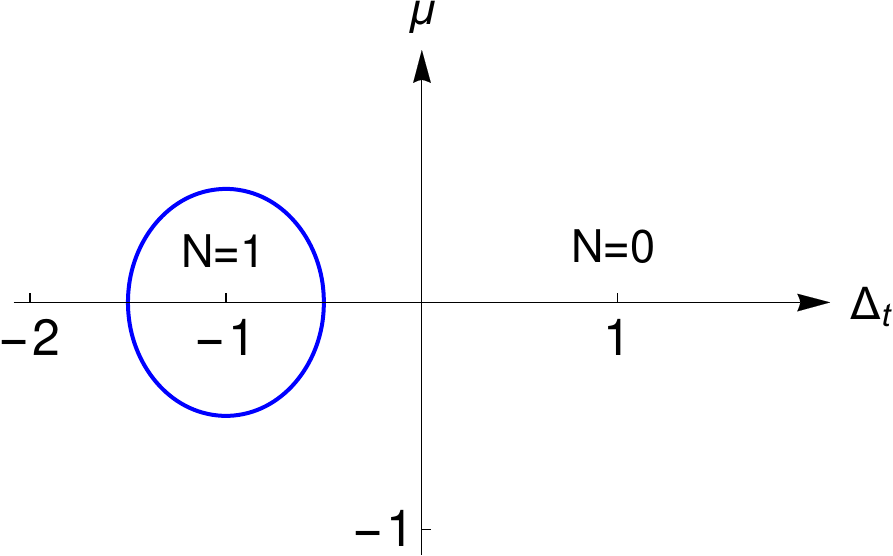}}
		\centerline{(a)}
	\end{minipage}
	\begin{minipage}{0.23\textwidth}
		\centerline{\includegraphics[width=1\textwidth]{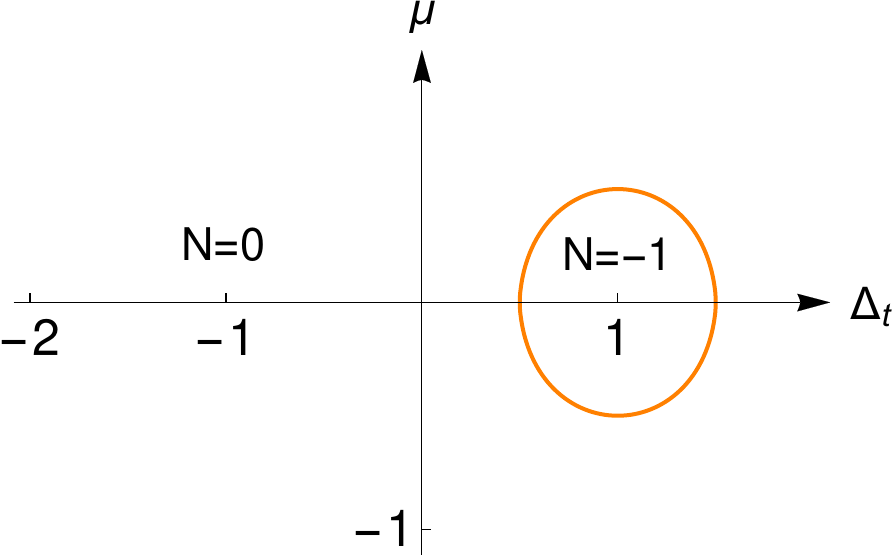}}
		\centerline{(b)}
	\end{minipage}
	\begin{minipage}{0.23\textwidth}
		\centerline{\includegraphics[width=1\textwidth]{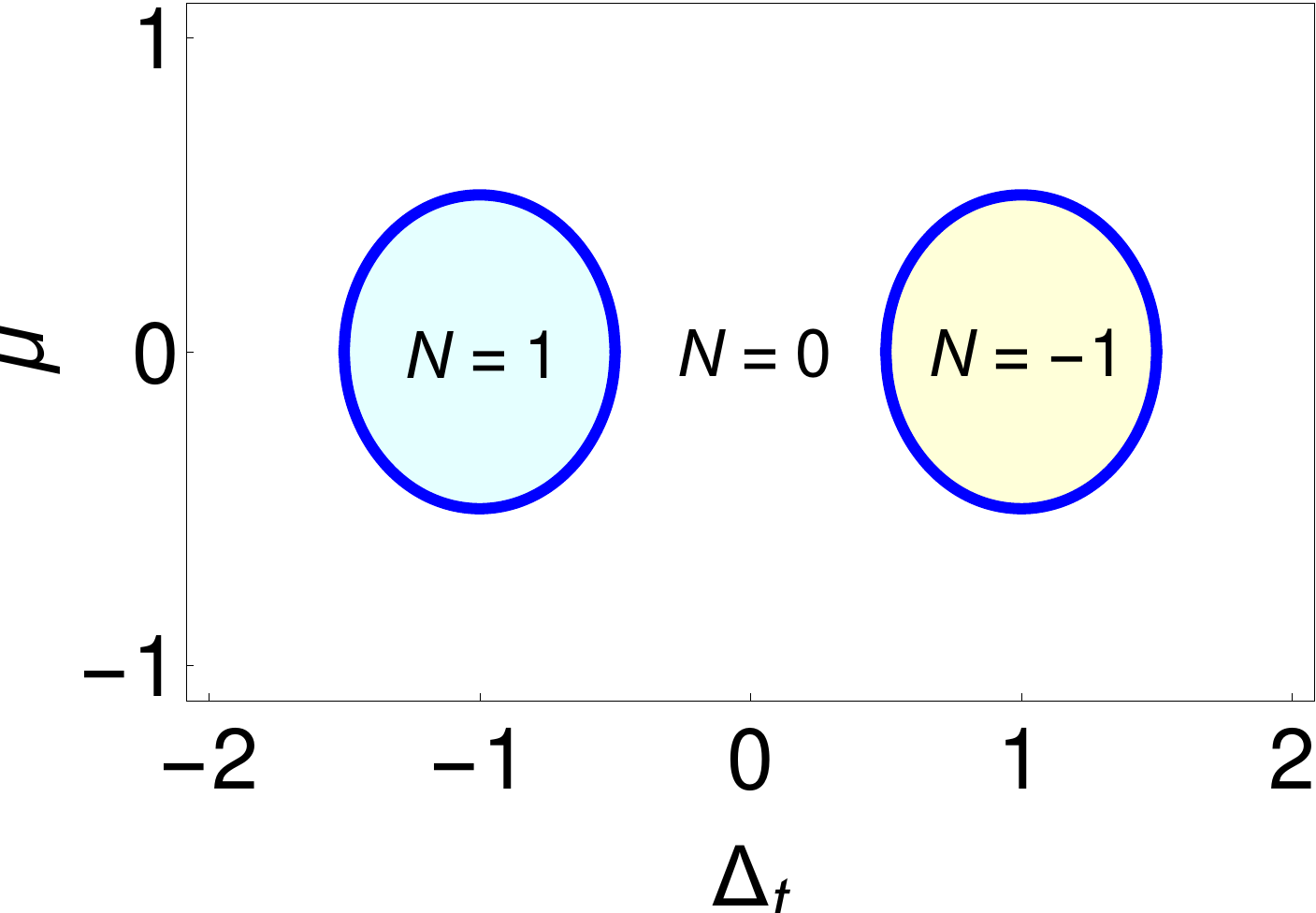}}
		\centerline{(c)}
	\end{minipage}}
	\caption{(Colour online)
	Topological phase diagrams of (a) $h_1$, (b) $h_2$, (c) $h$ in the $\Delta_t$-$\mu$ plane for $\lambda=0$, $\Delta_s=0$, $\Delta=1$, and $t=0.5$. }
		\label{fig2}	
\end{figure}

\begin{figure}
	\centerline{
	\begin{minipage}{0.23\textwidth}
		\centerline{\includegraphics[width=1\textwidth]{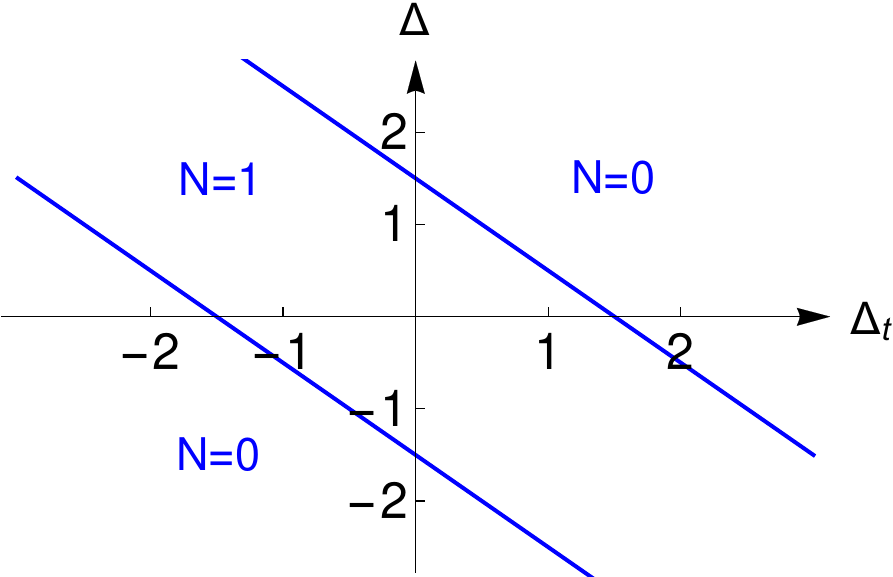}}
		\centerline{(a)}
	\end{minipage}
	\begin{minipage}{0.23\textwidth}
		\centerline{\includegraphics[width=1\textwidth]{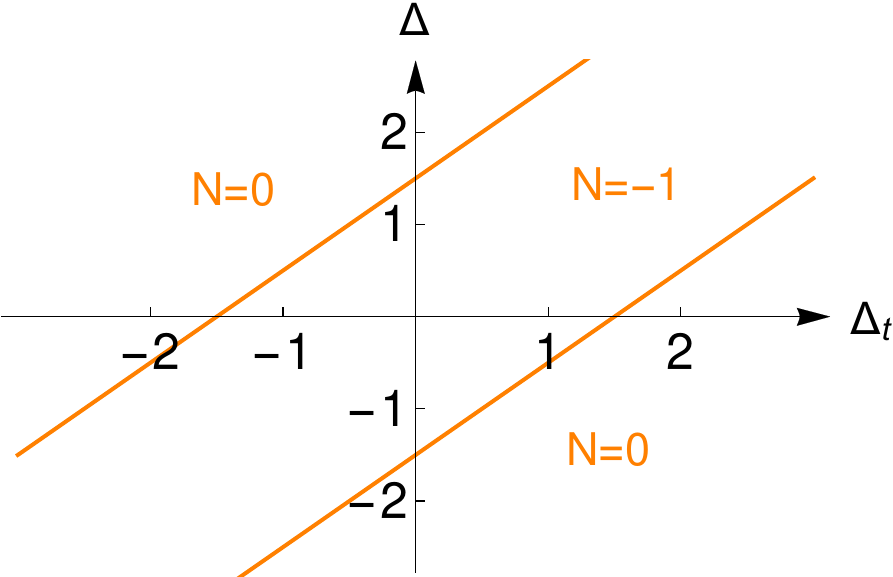}}
		\centerline{(b)}
	\end{minipage}
	\begin{minipage}{0.23\textwidth}
		\centerline{\includegraphics[width=1\textwidth]{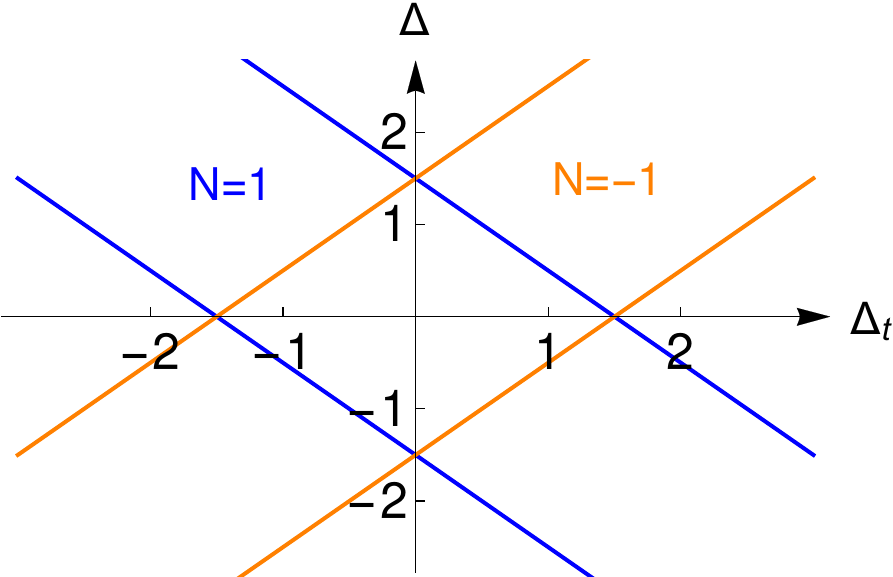}}
		\centerline{(c)}
	\end{minipage}}
	\caption{(Colour online)
		Topological phase diagrams of (a) $h_1$, (b) $h_2$, (c) $h$ in the $\Delta_t$-$\Delta$ plane for $\lambda=0$, $\Delta_s=0$, $\mu=0$, and $t=1.5$. }
	\label{figs4}	
\end{figure}

In this section, we focus on the $\mathbb{Z}\oplus\mathbb{Z}$ case and discuss the effect of $\mathcal{M}$-even term $\Delta_s$. As discussed in section 2, the topological phase boundary is determined by the gap closing at the $\Gamma$ point. The gap of equation~(\ref{h}) with $\mu=\lambda=0$ at $\Gamma$ point is
\begin{equation}
	M_{\kappa\eta\xi}=\kappa\Delta_t+\eta\sqrt{(t+\xi\Delta)^2+\Delta_s^2}, \label{E0}
\end{equation} 
with $\kappa,\eta,\xi=\pm$. The gaps at $\Gamma$ point for $h_1$ and $h_2$ are
\begin{equation}
M_{1\kappa\eta}=\kappa(t+\lambda)+\eta\sqrt{(\Delta+\Delta_t)^2+\mu^2}, \label{M1}
\end{equation} 
\begin{equation}
M_{2\kappa\eta}=\kappa(t-\lambda)+\eta\sqrt{(\Delta-\Delta_t)^2+\mu^2}. \label{M2}
\end{equation}
Here, the ones with $h_3$ and $h_4$ are
\begin{equation}
M_{3\kappa\eta}=\kappa\lambda+\eta\sqrt{(t-\kappa\mu)^2+\Delta_s^2}, \label{M3}
\end{equation}
\begin{equation}
M_{4\kappa\eta}=\kappa\lambda+\eta\sqrt{(t+\kappa\mu)^2+\Delta_s^2}. \label{M4}
\end{equation}
When $\lambda=0$ and $\mu=0$, the phase boundaries represented by equation~(\ref{E0}) are exactly the superposition of the phase boundaries described by equations~(\ref{M1},~\ref{M2}) or equations~(\ref{M3}, \ref{M4}), correspondingly.  First we consider the phase diagram of $h_1$ and $h_2$ when $\Delta_s=0$. Clearly, the phase diagram of $8\times 8$ BdG model Hamiltonian is a direct sum of the those of $h_1$ and $h_2$ (see figure~\ref{fig1}, figure~\ref{fig2} and figure~\ref{figs4}). This can also be confirmed in the structure of phase boundaries, namely, according to the gap function $M_{1\kappa\eta}$ and  $M_{2\kappa\eta}$, the critical lines of figures~\ref{figs4}~(a) and~\ref{figs4}~(b) are determined by $M_{1\kappa\eta}=0$ and $M_{2\kappa\eta}=0$, and the combination of these two sets of phase boundaries are identical to the ones from the gap function of $M_{\kappa\eta\xi}=0$. Furthermore, in the presence of the Zeeman term, because $h_1(\Delta,\lambda, \mu, q_2)=h_2(-\Delta,-\lambda, -\mu,  -q_2)$, there are symmetries in the phase diagrams. In general,  $N_1(\Delta_t,\lambda)=-N_2(-\Delta_t,-\lambda)$, where $N_1(N_2)$ is the Chern number that corresponds to $h_1(h_2)$~\cite{chern0,chern}. Therefore, by monotonously tuning $\lambda$, the total Chern number can change from $2$ to $-2$, see figure~\ref{fig1}~(c).


\begin{figure}[htp]
	\centerline{
	\begin{minipage}{0.23\textwidth}
		\centerline{\includegraphics[width=1\textwidth]{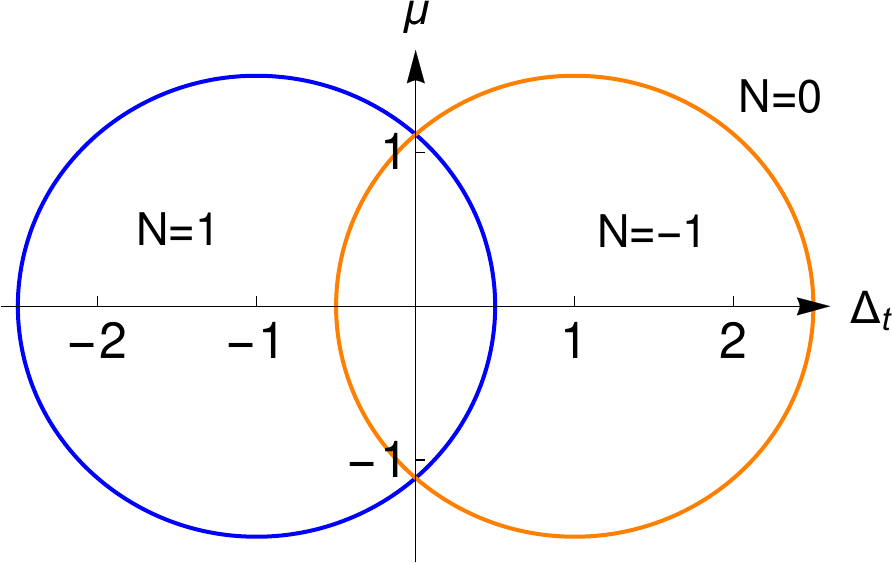}}
		\centerline{(a)}
	\end{minipage}
	\begin{minipage}{0.23\textwidth}
		\centerline{\includegraphics[width=1\textwidth]{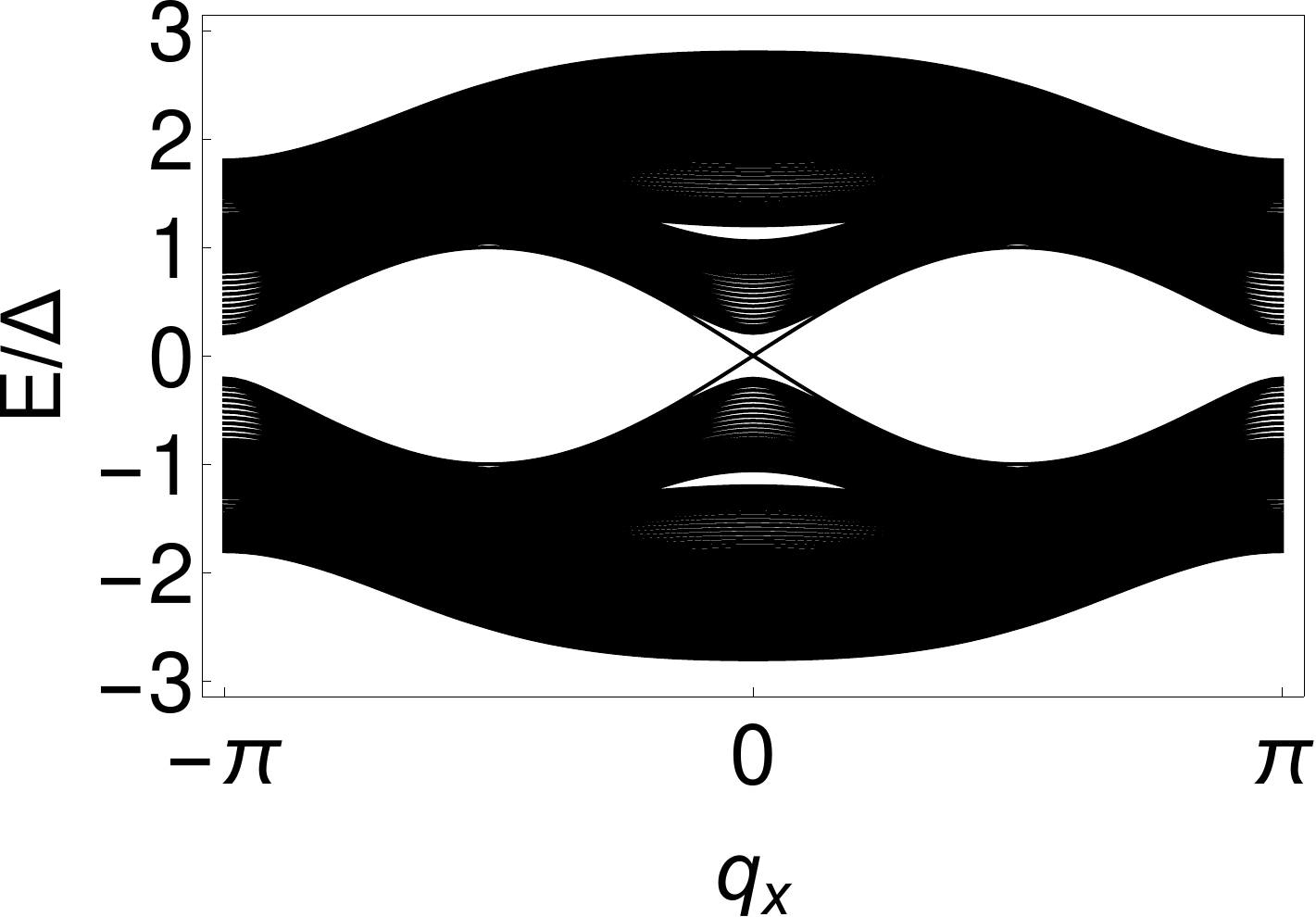}}
		\centerline{(b)}
	\end{minipage}
	\begin{minipage}{0.23\textwidth}
		\centerline{\includegraphics[width=1\textwidth]{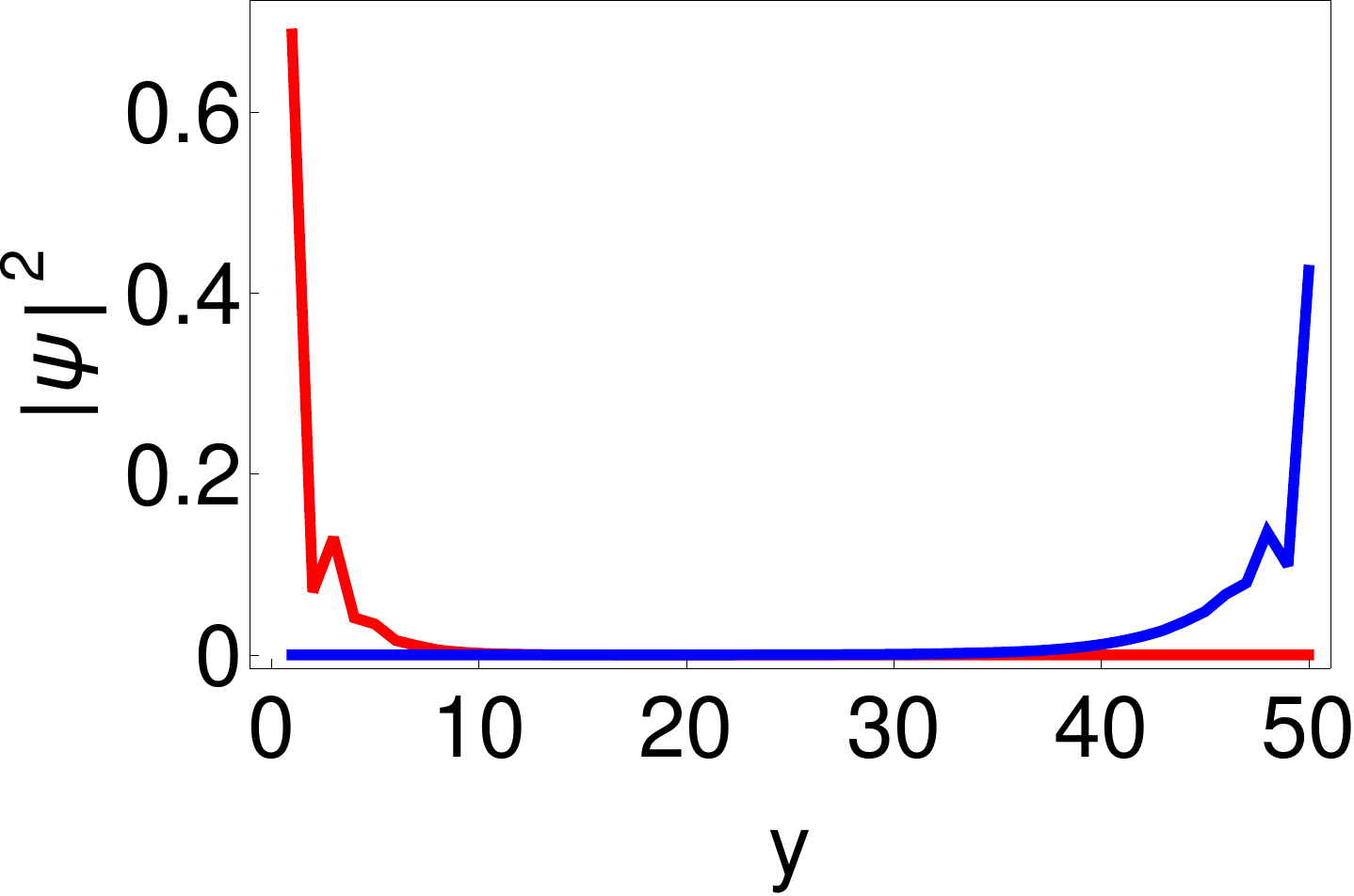}}
		\centerline{(c)}
	\end{minipage}}
	\caption{(Colour online)
		(a) Topological phase diagrams of $h_1$ and $h_2$  in the $\Delta_t$-$\mu$ plane with $\lambda=0$, $\Delta_s=0$, $\Delta=1$, and $t=1.5$. (b) The band structure of $h$ with 50-sites along the $y$-direction with $\lambda=\mu=\Delta_s=0$, $\Delta=1$, $t_0=t=1.5$,  $t_1=-0.5$ and $\Delta_t=0.3$. (c) The wave function distributions of the two edge states associated with positive slope at $q_x=0.5$. They localize at the opposite edges, which means that for each edge, there are two orthogonal chiral Majorana edge modes propagating along the opposite directions. }
	\label{fig3}	
\end{figure}

\begin{figure}[htp]
	\centerline{
	\begin{minipage}{0.35\textwidth}
		\centerline{\includegraphics[width=1\textwidth]{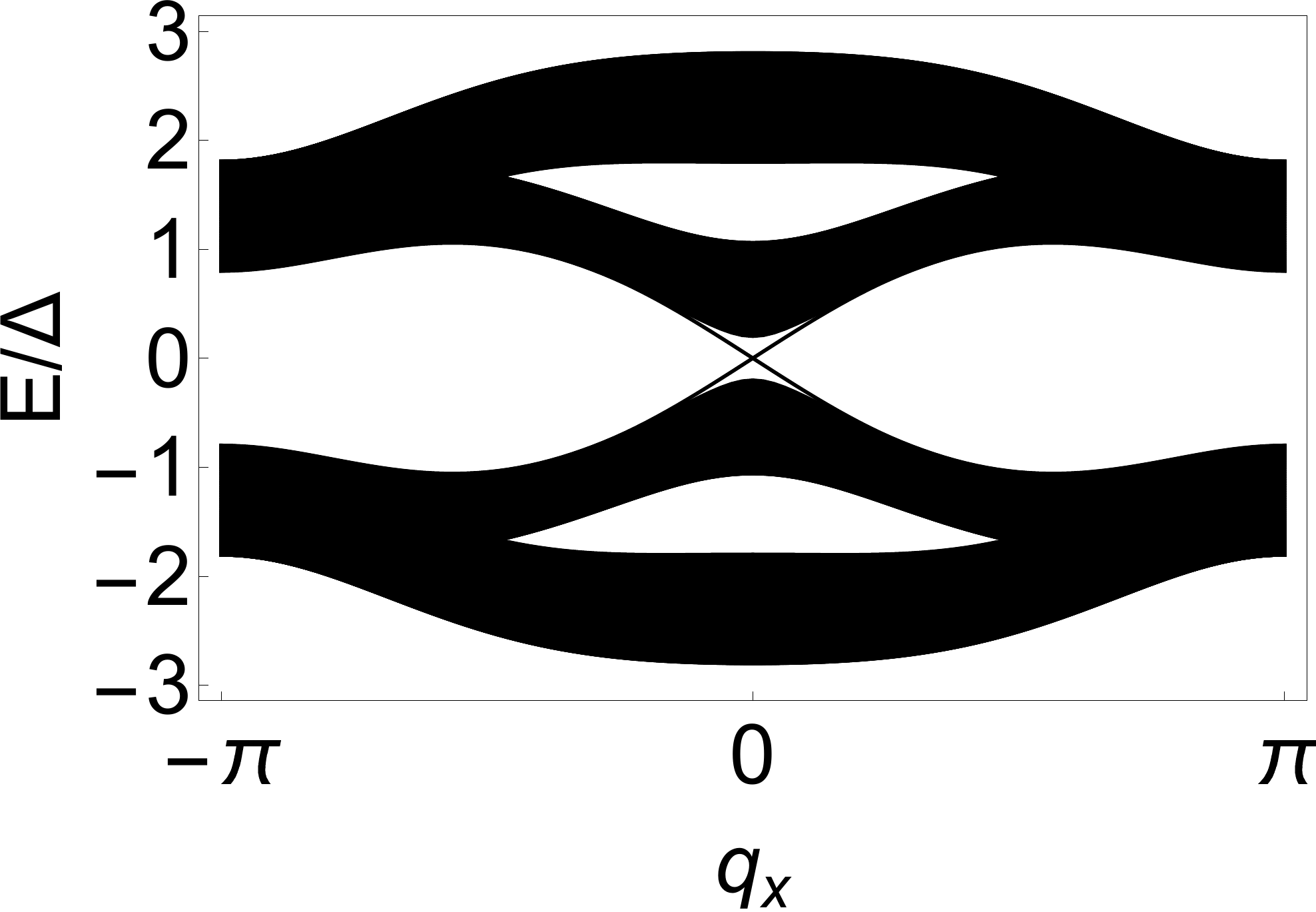}}
		\centerline{(a)}
	\end{minipage}
	\begin{minipage}{0.35\textwidth}
		\centerline{\includegraphics[width=1\textwidth]{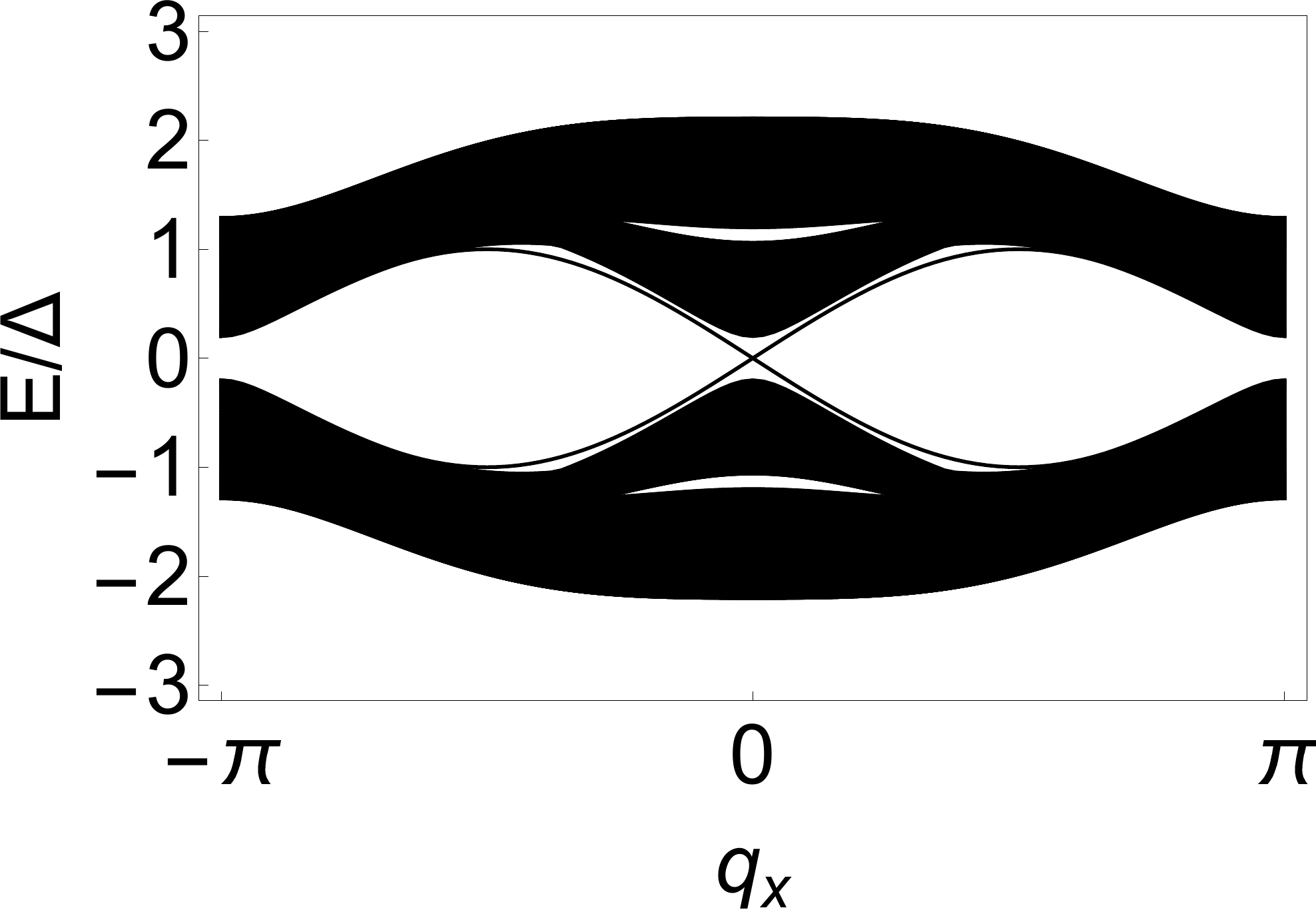}}
		\centerline{(b)}
	\end{minipage}}
	\centerline{
	\begin{minipage}{0.35\textwidth}
		\centerline{\includegraphics[width=1\textwidth]{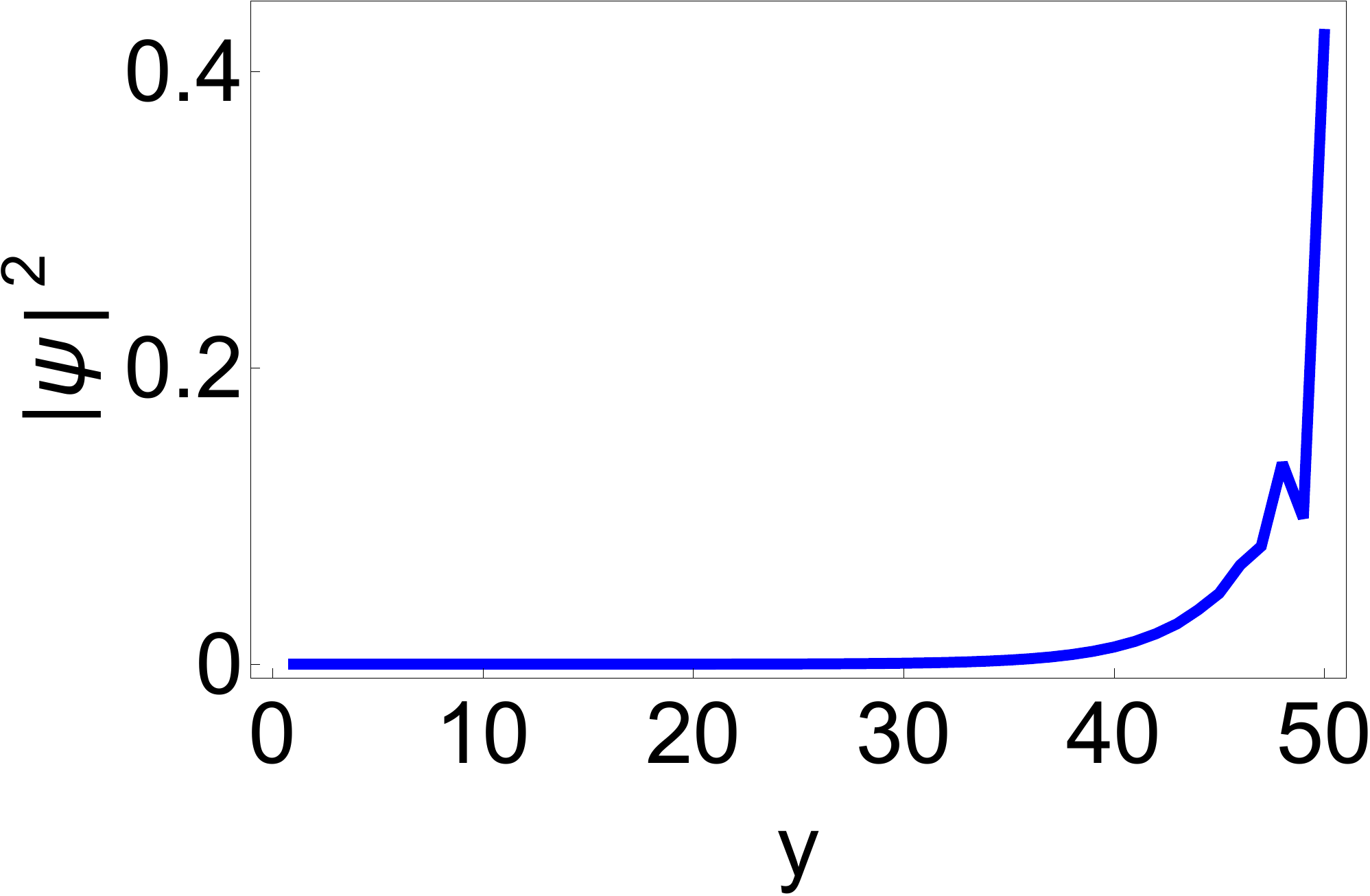}}
		\centerline{(c)}
	\end{minipage}
	\begin{minipage}{0.35\textwidth}
		\centerline{\includegraphics[width=1\textwidth]{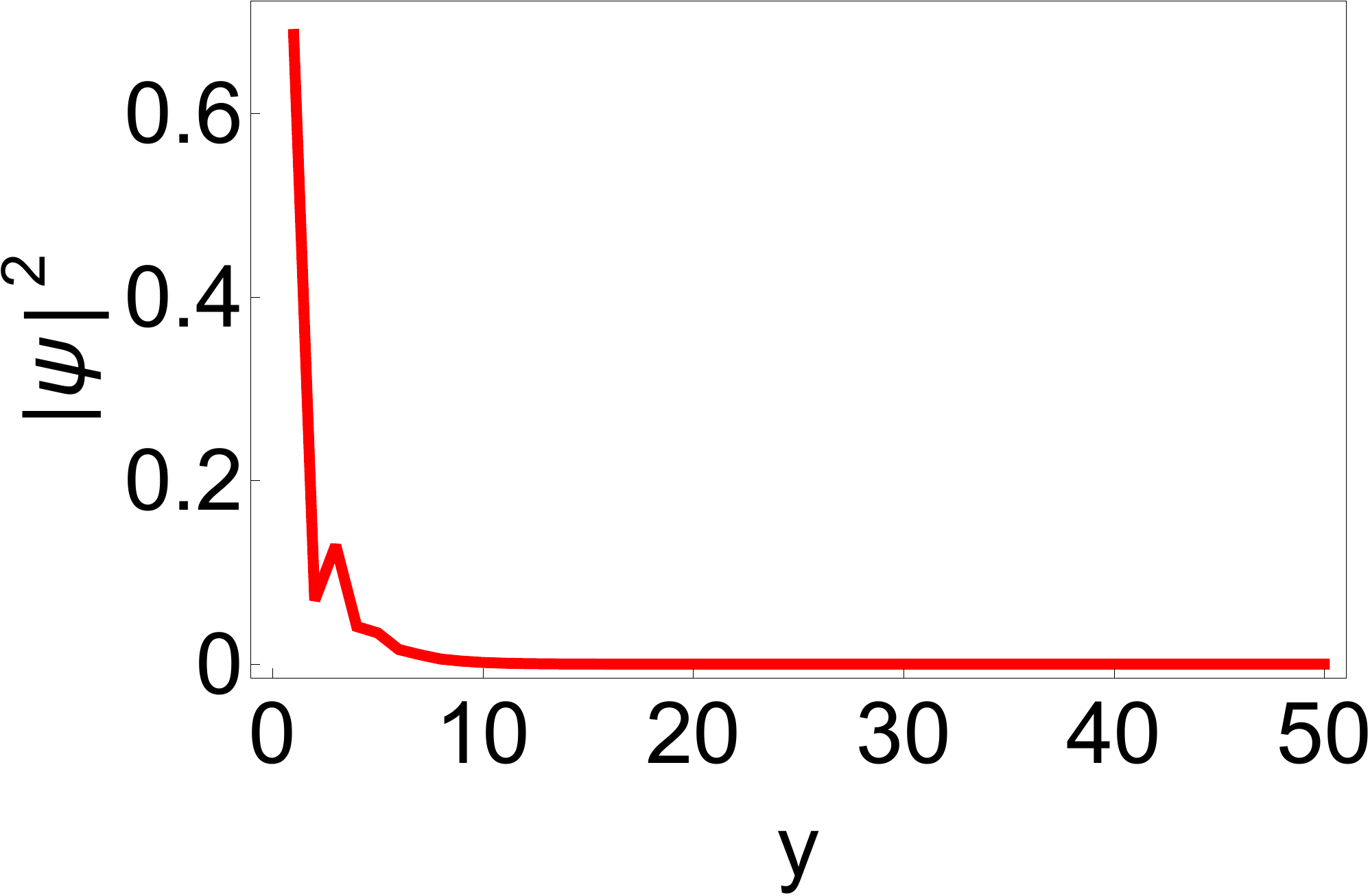}}
		\centerline{(d)}
	\end{minipage}}
	\centerline{
	\begin{minipage}{0.37\textwidth}
		\centerline{\includegraphics[width=1\textwidth]{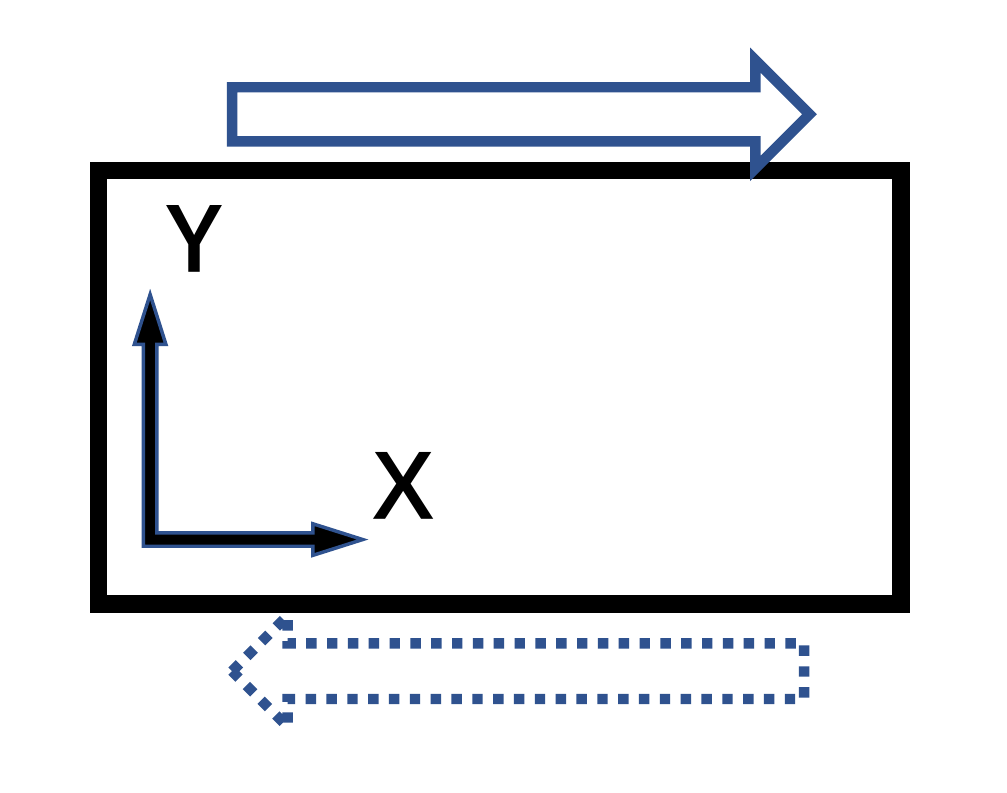}}
		\centerline{(e)}
	\end{minipage}
	\begin{minipage}{0.4\textwidth}
		\centerline{\includegraphics[width=1\textwidth]{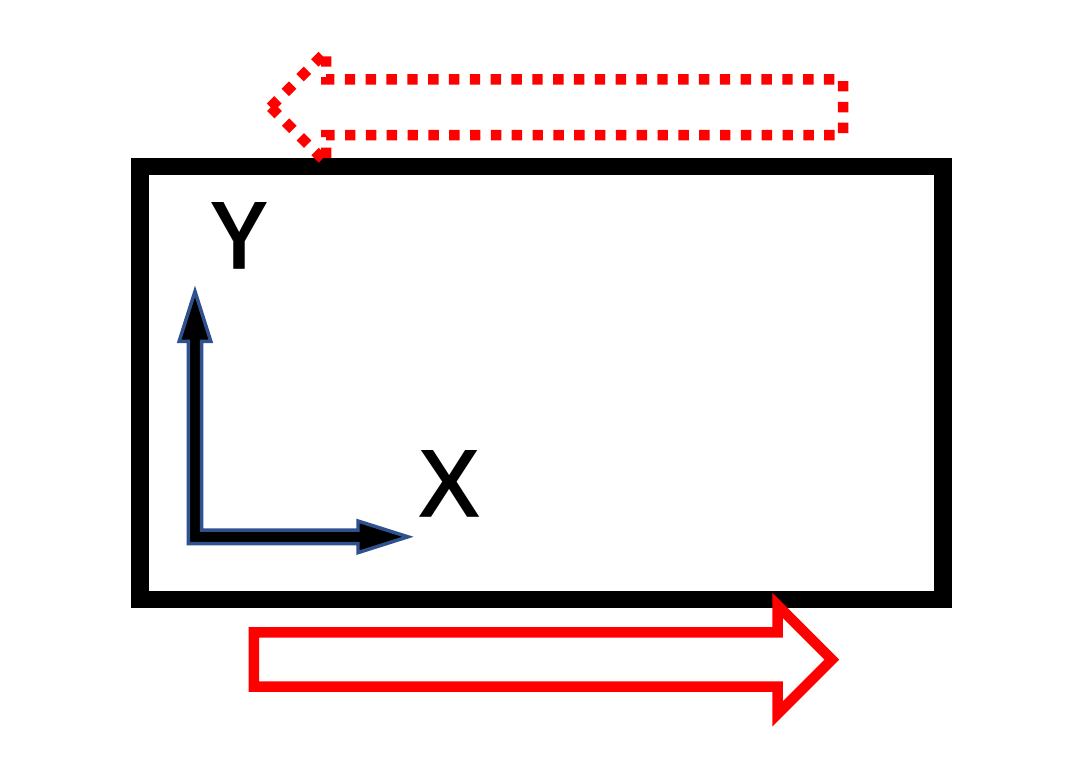}}
		\centerline{(f)}
	\end{minipage}}
	\caption{(Colour online)
	(a) and (b) are the band structure of $h_1$ and $h_2$, (c) and (d) are the spatial distribution of the edge states at $q_x=0.5$. The corresponding paremeters are the same as in figure~\ref{fig3}~(b) and~\ref{fig3}~(c). (e) and (f) are the schematic top views of the edge states corresponding to (c) and (d), the dashed arrows are the edges states with negative slope in the band structures in (a) and (b). } 
	\label{fig6}	
\end{figure}

Let us now consider the non-trivial case of $(N=1)\oplus(N=-1)$, as in the intersecting area between the two circles in figure~\ref{fig3}~(a). Though the total Chern number is $1+(-1)=0$, this phase is topologically non-trivial. We demonstrate this phase through a lattice model study by substituting $q_1\rightarrow\sin(q_1)$, $q_2\rightarrow\sin(q_2)$, and $t\rightarrow t_0+t_1[2-\cos(q_1)-\cos(q_2)]$. After the substitution, the Chern number  of each region in the lattice model can be calculated by~\cite{chern}
\begin{equation}
	c_n=\sum_ln_{12}(k_l),
\end{equation}
where $k_l$ denotes lattice points on the discrete Brillouin zone, $n_{12}$ is an integer value defined by the Berry's phase. For example, there are two edge states associated with a positive slope in figure~\ref{fig3}~(b), and one is localized on one edge and the other is on the other edge [see figure~\ref{fig3}~(c)], indicating the $(N=1)\oplus(N=-1)$ structure. Each of them is Majorana fermion since $h_1$ and $h_2$ are both particle-hole symmetric. These Majorana edge modes are gapless as long as the mirror symmetry is preserved. To further explain the $(N=1)\oplus(N=-1)$ structure with the edge state,  we draw the band structures and spatial distributions of edge states of $h_1$ and $h_2$ separately in figure~\ref{fig6}. By comparing figures~\ref{fig3} and~\ref{fig6}, the topological classification of  $\mathbb{Z}\oplus \mathbb{Z}$ is confirmed. 

\begin{figure}[htp]
		\centerline{\includegraphics[width=0.3\textwidth]{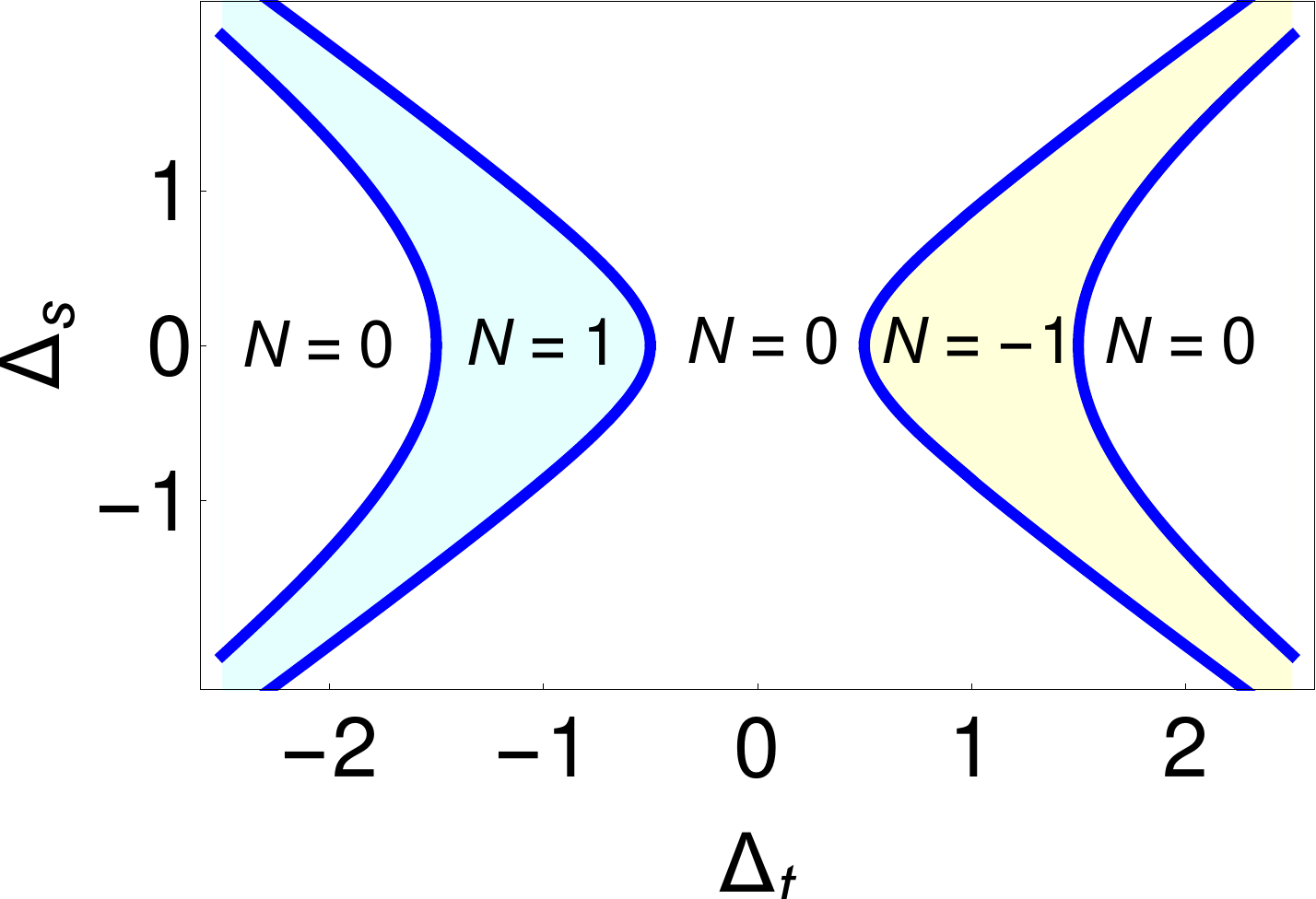}}
	\caption{(Colour online)
		Topological phase diagrams in of $h$ in $\Delta_t$-$\Delta_s$ plane for $\mu=0$, $\Delta=1$, and $t=0.5$. }
	\label{fig4}	
\end{figure}

Now, we are ready to discuss the effects of $\Delta_s$. From figure~\ref{fig4}, one can observe that, the $N=1$ and $N=-1$ areas correspond to the non-trivial phase of $h_1$ and $h_2$, respectively, because they are continuously connected to $\Delta_s=0$ case. Though $\Delta_s$ breaks $M_-$ symmetry, the topological gap remains open when $\Delta_s$ is small, and within a proper range, $\Delta_s$ can drive the topological trivial phase into a non-trivial one (see the left-hand and right-hand $N=0$ areas in figure~\ref{fig4}).  

\section{Effects of complex pairings}

In this section, we will consider the effects of complex pairings, namely, there can be a relative phase between $\Delta$ and $\Delta_t$. To be more specific, we substitute $\Delta\rightarrow\Delta_1,\Delta_2$, where they are the strengths of the intra-surface pairings on each surface, and real. We also replace $\Delta_t=\Delta_{t1}+\ri\Delta_{t2}$, with $\Delta_{t1}$ and $\Delta_{t2}$ being real. For the mirror symmetry, we still use $M_-=-\ri\sigma_z \chi_x \tau_0$, and our Hamiltonian is
\begin{small}
\begin{eqnarray}
&&U_- h(\textbf{q})U_-^{-1}\nonumber \\
&=&
\left(
\begin{array}{cccccccc}
	-t+\mu  & \alpha +\Delta _t{}^* & 0 & q_1+\ri q_2 & 0 & 0 & 0 & \beta  \\
	\alpha +\Delta _t & -t-\mu  & q_1+\ri q_2 & 0 & 0 & 0 & \beta  & 0 \\
	0 & q_1-\ri q_2 & t-\mu  & -\alpha -\Delta _t & 0 & -\beta  & 0 & 0 \\
	q_1-\ri q_2 & 0 & -\alpha -\Delta _t{}^* & t+\mu  & -\beta  & 0 & 0 & 0 \\
	0 & 0 & 0 & -\beta  & -t-\mu  & -\alpha +\Delta _t & 0 & q_1-\ri q_2 \\
	0 & 0 & -\beta  & 0 & -\alpha +\Delta _t{}^* & -t+\mu  & q_1-\ri q_2 & 0 \\
	0 & \beta  & 0 & 0 & 0 & q_1+\ri q_2 & t+\mu  & \alpha -\Delta _t{}^* \\
	\beta  & 0 & 0 & 0 & q_1+\ri q_2 & 0 & \alpha -\Delta _t & t-\mu  \\
\end{array}
\right),
\end{eqnarray}
\end{small}
where we have set $\Delta_s=\lambda=0$, $\alpha=(\Delta_1+\Delta_2)/2$, and $\beta=(\Delta_1-\Delta_2)/2$. Clearly, $\alpha$, $\Delta_{t1}$, and $\Delta_{t2}$ terms preserve the $\mathcal{M}_-$ symmetry, and they are $\mathcal{M}$-odd, while $\beta$ term breaks the  $\mathcal{M}_-$ symmetry and it is $\mathcal{M}$-even. 

\begin{figure}
 	\centerline{
 	\begin{minipage}{0.3\textwidth}
 		\centerline{\includegraphics[width=1\textwidth]{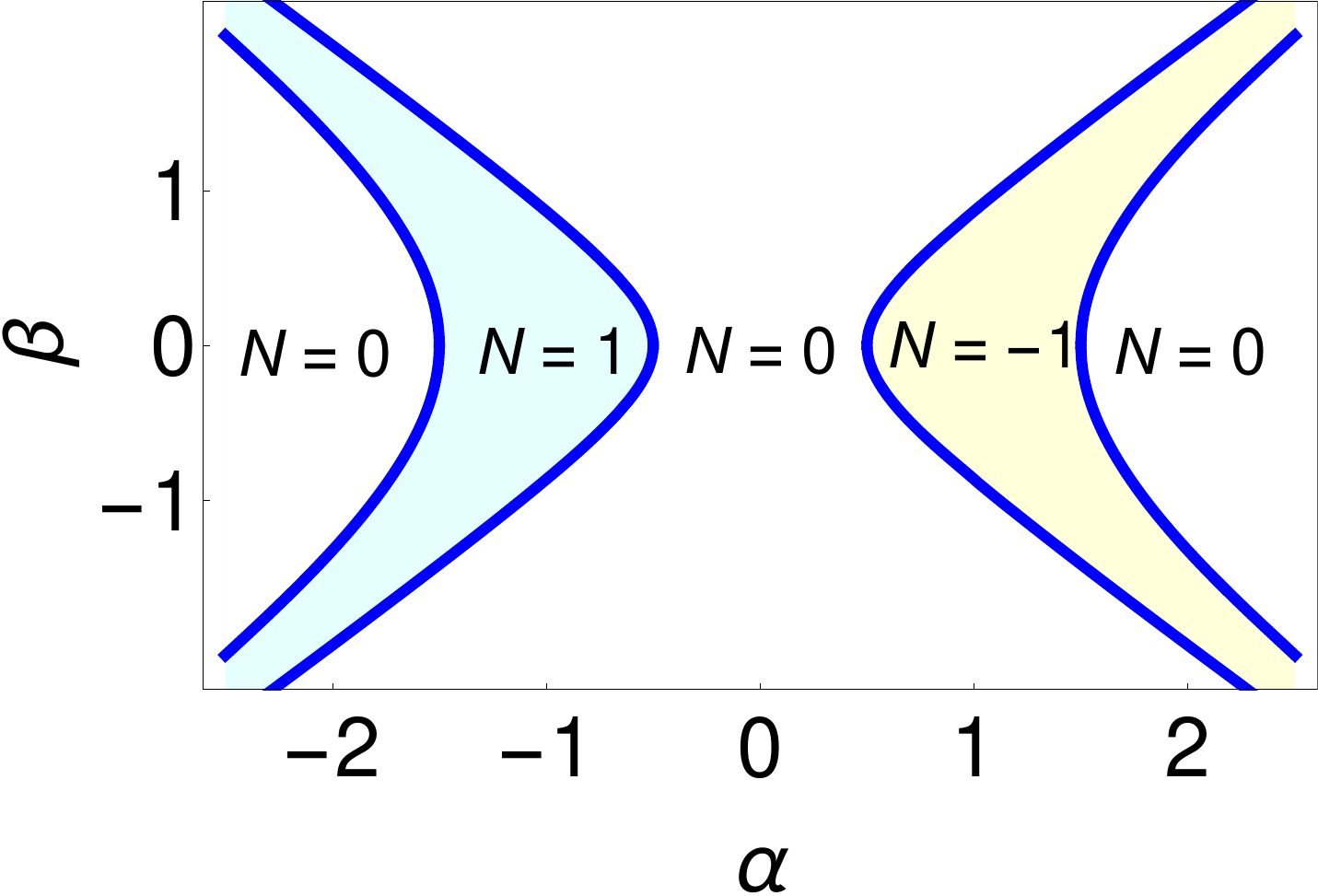}}
 		\centerline{(a)}
 	\end{minipage}
 	\begin{minipage}{0.3\textwidth}
 		\centerline{\includegraphics[width=1\textwidth]{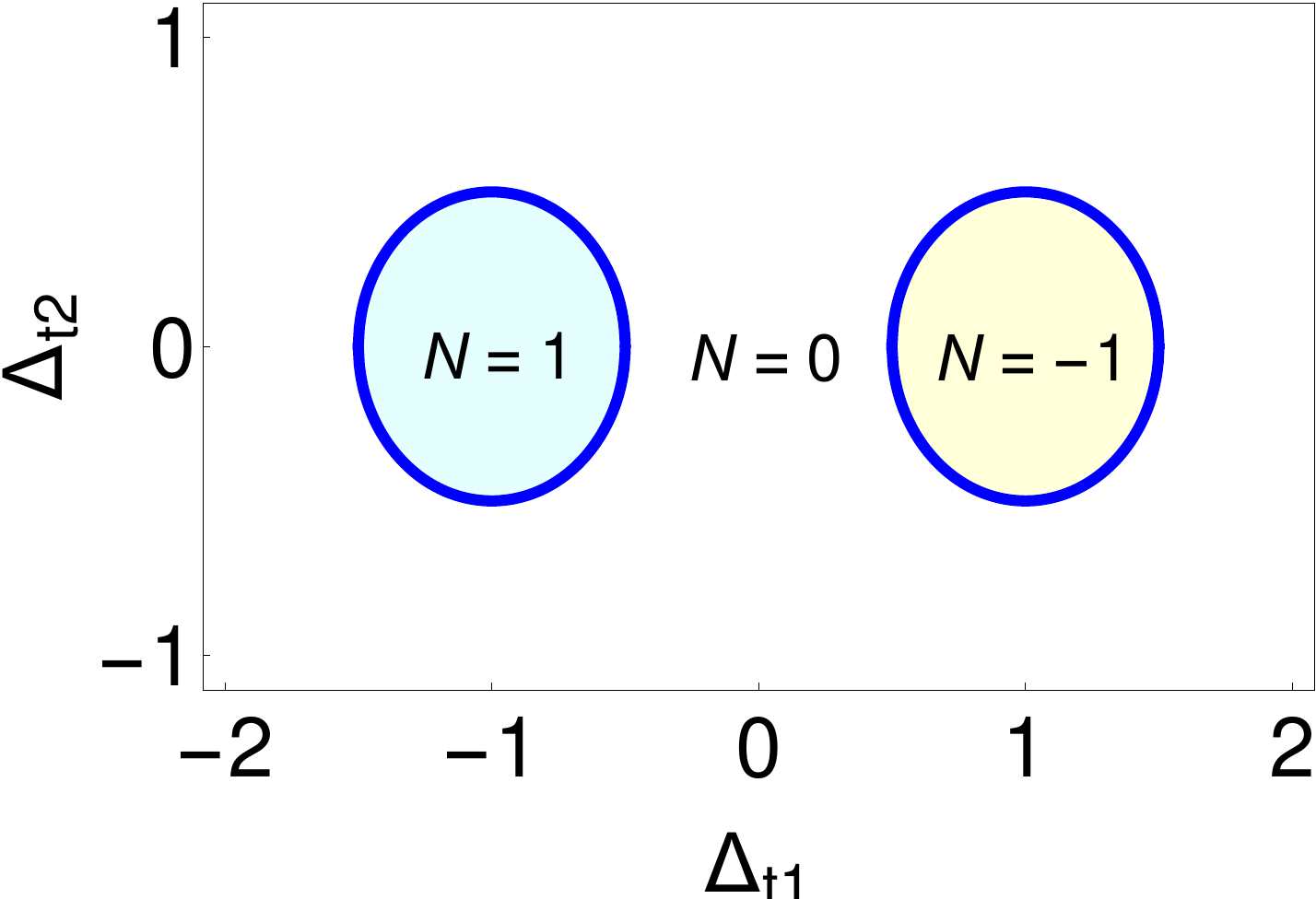}}
 		\centerline{(b)}
 	\end{minipage}}
 	\caption{(Colour online)
 		Topological phase diagrams of $h$ in (a) the $\alpha-\beta$ plane for $\mu=0$, $t=0.5$, and $\Delta_t=1$; (b) the $\Delta_{t1}-\Delta_{t2}$ plane for $\mu=0$, $t=0.5$, $\alpha=1$ and $\beta=0$.  }
 	\label{fig7}	
 \end{figure}

We draw the phase diagrams in the $\alpha-\beta$ plane and $\Delta_{t1}-\Delta_{t2}$ plane in figure~\ref{fig7}. By comparing figure~\ref{fig7} with figure~\ref{fig2}~(c) and figure~\ref{fig4}, we conclude that the difference in $\Delta_1$ and $\Delta_2$, namely, $\beta$ has a similar effect as the $\mathcal{M}_-$ breaking term $\Delta_s$, and the imaginary part of $\Delta_t$, namely, $\Delta_{t2}$ behaves as the chemical potential~$\mu$.  
 
\section{Conclusion}

In this paper, we present a detailed study on the topological superconductivity near the $\Gamma$ point of the topological surface states on the top and bottom surfaces. Without external magnetic field, we find a novel mechanism for spontaneous time-reversal symmetry breaking in such systems. Because of the emergent mirror symmetry between the two topological Dirac surface states of the thin film of a topological insulator, there are only two momentum independent pairing terms favored by the mirror symmetry, namely, one is the intra-surface spin singlet pairing $\Delta$, and the other is the inter-surface spin triplet pairing $\Delta_t$. When both of the pairings and the inter-surface hopping are non-zero, the time reversal symmetry is broken spontaneously. Here, because of the mirror symmetry, the topological classification is $\mathbb{Z}\oplus\mathbb{Z}$. The non-trivial topological band structure of the BdG Hamiltonian with non-zero bulk Chern number provides stable chiral Majorana edge modes propagating along the edge, which may pave the way for realizing topological quantum computation based on Majorana fermions~\cite{luo2}. The models we discussed are realizable in the thin films of iron-based topological superconductors such as FeTeSe with reachable experimental techniques. The chiral Majorana modes could be detected by STM similar to the case of the Majorana zero modes~\cite{lkong}, or  by the thermo-conductivity. We also discuss the effects of a relative phase between $\Delta$ and $\Delta_t$. The difference in the intra-surface pairing on each surface breaks the mirror symmetry, while the imaginary part of $\Delta_t$ preserves the mirror symmetry.

\section*{Acknowledgements}
{We thank Yue Yu and Ziqiang Wang for helpful discussions. This work is supported by NNSF of China with No. 11804223.}

\ukrainianpart

	\title{Фазові діаграми надпровідних топологічних поверхневих станів}

	\author{В.~Х. Жао, Л.~Л. Динь, Б.~В. Жоу, Й.~І. Ву, І. Бай, З.~І. Ман, С. Луо\orcid{0000-0001-9124-6409}}
	\address{Факультет фізики, науковий коледж, Шанхайський університет науки і технології, Шанхай 200093, КНР}

\makeukrtitle

	\begin{abstract}
	У цій статті представлено детальне дослідження фазових діаграм надпровідних топологічних поверхневих станів. Особливу увагу приділено взаємозв'язку між кристалічною симетрією та топологією ефективного гамільтоніана Боголюбова-де Жена. Показано, шо для кінематичного гамільтоніана нормального стану $4\times 4$ можна визнaчити дзеркальну симетрію $\mathcal{M}$, в той час як для $\mathcal{M}$-непарних спарювань $8\times8$ гамільтоніан Боголюбова-де Жена класифікується як $\mathbb{Z}\oplus\mathbb{Z}$, а симетрія оборотності часу порушується за своєю суттю. Топологічна нетривіальна фаза може підтримувати хіральні майоранівські крайові моди та може бути реалізована у тонких плівках надпровідників на основі заліза, таких як FeSeTe.

	\keywords симетрія, фазові діаграми, топологія, поверхневі стани
	
	\end{abstract}

\lastpage
\end{document}